\documentclass{aa}

\usepackage[utf8]{inputenc}
\usepackage{graphicx}
\usepackage{txfonts}

\usepackage[usenames,dvipsnames,svgnames,table]{xcolor}
\definecolor{cadmiumgreen}{rgb}{0.0, 0.7, 0.2}
\usepackage{hyperref}
\hypersetup{
  colorlinks = true,
  citecolor = cadmiumgreen
  }

\usepackage{natbib}
\usepackage{numdef}
\usepackage{multirow}
\usepackage{color,soul}
\usepackage{balance}

\newcommand{\np}{\ensuremath{N_p}\xspace}
\newcommand{\rhk}{$R'_{\rm HK}$\xspace}

\newcommand{\Msun}{$M_\odot$\xspace}

\newcommand{\Mearth}{$M_{\oplus}$\xspace}
\newcommand{\Rsun}{$R_\odot$\xspace}

\newcommand{\ica}{$\rm I_{\rm Ca\,II}$\xspace}
\newcommand{\iha}{$\rm I_{\rm H\alpha}$\xspace}
\newcommand{\ina}{$\rm I_{\rm Na\,I}$\xspace}
\newcommand{\fig}[1]{Fig.\,\ref{#1}\xspace}
\newcommand{\TESS}{\emph{TESS}\xspace}

\DeclareMathOperator{\cms}{cm\!\cdot\!s^{-1}} 
\DeclareMathOperator{\ms}{\,m\!\cdot\!s^{-1}} 
\DeclareMathOperator{\kms}{\,km\!\cdot\!s^{-1}} 
\DeclareMathOperator{\msy}{\,m\!\cdot\!s^{-1}yr^{\,-1}} 

\begin{document} 

  \title{Decoding the radial velocity variations of HD41248 with ESPRESSO}
  \titlerunning{Decoding the radial velocity variations of HD41248 with ESPRESSO}

  \author{
J.~P.~Faria                    \inst{1}
\and
V.~Adibekyan                   \inst{1}
\and
E.~M.~Amazo-Gómez              \inst{2,3}
\and
S.~C.~C.~Barros                \inst{1}
\and
J.~D.~Camacho                  \inst{1,4}
\and
O.~Demangeon                   \inst{1}
\and
P.~Figueira                    \inst{5,1}
\and
A.~Mortier                     \inst{6}
\and
M.~Oshagh                      \inst{3,1}
\and
F.~Pepe                        \inst{7}
\and
N.~C.~Santos                   \inst{1,4}
\and
J.~Gomes~da~Silva              \inst{1}
\and
A.~R.~Costa~Silva              \inst{1,8}
\and
S.~G.~Sousa                    \inst{1}
\and
S.~Ulmer-Moll                  \inst{1,4}
\and
P.~T.~P.~Viana                 \inst{1}
}
\institute{
Instituto de Astrofísica e Ciências do Espaço, Universidade do Porto, CAUP, Rua das Estrelas, 4150-762 Porto, Portugal
\and
Max-Planck-Institut für Sonnensystemforschung, Justus-von-Liebig-Weg 3, 37077 Göttingen, Germany
\and
Institut für Astrophysik, Georg-August-Universität, Friedrich-Hund-Platz 1, D-37077, Göttingen, Germany
\and
Departamento de Física e Astronomia, Faculdade de Ciências, Universidade do Porto, Rua Campo Alegre, 4169-007 Porto, Portugal
\and
European Southern Observatory, Alonso de Cordova 3107, Vitacura, Santiago, Chile
\and
Astrophysics Group, Cavendish Laboratory, University of Cambridge, J.J. Thomson Avenue, Cambridge CB3 0HE, UK
\and
Observatoire Astronomique de l’Université de Genève, 51 chemin des Maillettes, 1290, Versoix, Switzerland
\and
University\,of\,Hertfordshire,\,School\,of\,Physics,\,Astronomy\,and\,Mathematics,\,College\,Lane\,Campus,\,Hatfield,\,Hertfordshire,\,AL10\,9AB,\,UK.
}

  \date{Received July 26, 2019; accepted November 21, 2019}

  \def\target{HD\,41248\xspace}
  \def\targetothername{HIP\,28460\xspace}

  \abstract
  {Twenty-four years after the discoveries of the first exoplanets, the radial-velocity
  (RV) method is still one of the most productive techniques to detect and
  confirm exoplanets. But stellar magnetic activity can induce RV variations
  large enough to make it difficult to disentangle planet signals from the
  stellar noise. In this context, \target is an interesting planet-host
  candidate, with RV observations plagued by activity-induced signals.}
  %
  {We report on ESPRESSO observations of \target and analyse them together with
  previous observations from HARPS with the goal of evaluating the presence of
  orbiting planets.}
  %
  {Using different noise models within a general Bayesian framework designed for
  planet detection in RV data, we test the significance of the various signals
  present in the \target dataset. We use Gaussian processes as well as a
  first-order moving average component to try to correct for activity-induced
  signals. At the same time, we analyse photometry from the \TESS mission,
  searching for transits and rotational modulation in the light curve.}
  %
  {The number of significantly detected Keplerian signals depends on the noise
  model employed, which can range from 0 with the Gaussian process model to 3 with a
  white noise model. We find that the Gaussian process alone can explain the RV
  data while allowing for the stellar rotation period and active region evolution
  timescale to be constrained. The rotation period estimated from the RVs agrees
  with the value determined from the \TESS light curve.}
  {Based on the data that is currently available, we conclude that the RV variations of
  \target can be explained by stellar activity (using the Gaussian process
  model) in line with the evidence from activity indicators and the \TESS
  photometry.}


  \maketitle

\section{Introduction}
\label{sec:intro}

  After a substantial growth in the number of exoplanet discoveries, we are now
  close to reaching the required precision for the detection of Earth-like
  planets around stars like the Sun. The radial-velocity (RV) method, in
  particular, has been instrumental for many planet detections and for measuring
  precise masses for transiting planets. However, detecting low-mass planets in
  RV time series is not an easy task due to contaminating signals originating
  from the stars themselves or instrumental effects \citep[e.g.][]{Fischer2016}.

  Stellar activity has long been recognised as an important source of RV
  variations \cite[e.g.][]{Saar1997,Santos2000}. The presence of active regions
  (spots and faculae) on the stellar surface can induce RV signals of several
  $\ms$, which can sometimes mimic planet signals \citep[e.g.][]{Figueira2010}
  or make it difficult to accurately characterise the planets' orbital
  parameters and masses. Several activity indicators (obtained from the same
  spectra where RVs are measured or from auxiliary data) have helped in
  disentangling planetary and activity signals in some favourable cases
  \citep[e.g.][]{Queloz2001,Melo2007}. Another approach is to develop
  sophisticated statistical models of stellar noise, which can efficiently
  separate the signals coming from different sources
  \citep[e.g.][]{Rajpaul2015}. 

  The metal-poor star \object{\target}$\!$ represents an interesting case in
  this respect. Using RV data from HARPS, \citet{Jenkins2013} announced the
  discovery of two super-Earths orbiting this star. The orbital periods of the
  proposed planets, around $\sim$18 and $\sim$25 days, would place them close to
  a 7:5 mean motion resonance, with implications for the formation history of
  this system. Later, \cite{Santos2014} analysed a much larger data set of HARPS
  RVs and could not confirm the planetary origin of the signals. The 25-day
  signal was also present in stellar activity indices, and the 18-day signal
  could not be recovered with significance in the new data. 
  
  A third paper from \citet{Jenkins2014} presents the analysis of the same
  extended HARPS data set and concludes, as before, that the two signals were
  induced by orbiting planets. The main differences between these works
  regard the model used to fit the RV data set: \citet{Jenkins2014} used a
  moving average term (MA, see Section \ref{sec:4}) to account for correlated
  noise and linear correlations with some activity indicators, while
  \citet{Santos2014} relied on white-noise models.

  More recently, \citet{Feng2017b} explored the full set of HARPS data again,
  concluding that the previously claimed $\sim$25-day signal is probably caused
  by `a combination of planetary perturbation and stellar rotation' while the
  $\sim$18-day signal is no longer significant and it is only an alias of a $\sim$13-day
  signal. This new signal was attributed to a planet candidate with a mass of
  7.08 \Mearth. \citet{Feng2017b} used a method based on periodograms which can
  take into account MA terms and linear correlations with activity indicators
  but only sinusoidal planetary signals.

  In this work, we present new ESPRESSO observations of \target and analyse the
  combined HARPS+ESPRESSO dataset. The paper is organised as follows: Section
  \ref{sec:2-observations}, presents the HARPS and ESPRESSO observations of
  \target, as well as the data reduction steps performed for each instrument. We
  also analysed TESS observations from the first ten sectors of the mission. In
  Section \ref{sec:3}, the stellar properties are derived from the observed
  spectra. Section \ref{sec:4} presents the methods to analyse the RVs, and
  Section \ref{sec:5} shows the results. We discuss our results and conclude in
  Sections \ref{sec:6} and \ref{sec:7}.

\section{Observations}
\label{sec:2-observations}

  In this section, we describe the observations from HARPS, ESPRESSO, and TESS,
  as well as the data reduction steps for each instrument.

  \subsection{HARPS}

  Between October 2003 and January 2014, \target was observed a total of
  230
times with the HARPS spectrograph%
  \footnote{Observations were obtained as part of the ESO programs with IDs
  \href{http://archive.eso.org/wdb/wdb/eso/abstract/query?progid=072.C-0488}{072.C-0488}, \href{http://archive.eso.org/wdb/wdb/eso/abstract/query?progid=082.C-0212}{082.C-0212}, \href{http://archive.eso.org/wdb/wdb/eso/abstract/query?progid=085.C-0063}{085.C-0063}, \href{http://archive.eso.org/wdb/wdb/eso/abstract/query?progid=086.C-0284}{086.C-0284}, \href{http://archive.eso.org/wdb/wdb/eso/abstract/query?progid=090.C-0421}{090.C-0421}, \href{http://archive.eso.org/wdb/wdb/eso/abstract/query?progid=098.C-0366}{098.C-0366}, \href{http://archive.eso.org/wdb/wdb/eso/abstract/query?progid=183.C-0972}{183.C-0972}, and \href{http://archive.eso.org/wdb/wdb/eso/abstract/query?progid=190.C-0027}{190.C-0027}.} %
  at the ESO 3.6\,m telescope in La Silla \citep{Mayor2003}. Before March 2013,
  simultaneous wavelength calibration was achieved using a ThAr lamp, whereas
  after that date, a Fabry-Perot etalon was used. Within the spectral order of
  50 (around 5500~\AA), the average signal-to-noise ratio (S/N) of the HARPS
  spectra is 86, with values ranging from 19 to 142.

  The very last HARPS spectrum was taken after a major intervention in the
  instrument which replaced the circular optical fibres for ones with an
  octagonal section, providing increased scrambling \citep{LoCurto2015}. This
  change introduced an RV offset of several $\ms$. With only one measurement it
  is very difficult to characterise the offset and, thus, this last point (with
  BJD$\simeq$2457766) was excluded from the analysis.

  \citet{Santos2014} also identified two other HARPS spectra with very low S/N
  or with an abnormal blue-to-red flux ratio. This can occur on nights when the
  transmission was affected by bad weather (e.g. due to cirrus). These two
  spectra (at BJD$\simeq$2455304
and BJD$\simeq$2456409) will also be excluded from the discussion below. Nevertheless, we
  compared the results of our analysis, including and excluding these two spectra,
  and found no significant differences.

  After these cuts, we have a total of 227
HARPS observations, with a
  time span of more than 10 years. The RVs were extracted from these spectra with
  the cross-correlation technique, as implemented by the HARPS pipeline. This
  method cross-correlates the observed spectrum with a weighted mask (in the
  case of \target, the mask is optimised for a G2 dwarf), giving rise to the
  cross-correlation function (CCF). The RV is calculated by fitting a Gaussian
  function to the CCF. The full width at half maximum (FWHM), depth, and
  bisector of the CCF were also calculated, as they are often used as stellar
  activity indicators \citep[e.g.][]{Queloz2001,Figueira2013,Lanza2018}.

  The uncertainties on the RVs include contributions from photon noise,
  wavelength calibration noise, and the uncertainty in the measurement of the
  instrumental drift \citep{Bouchy2001}. The full set of HARPS RVs and
  associated uncertainties are shown in \fig{fig:RVs} and listed in Table
  \ref{tab:rvs}. The average RV error of the 227
observations%
  \footnote{Note that we do not bin the observations per night.} %
  is 1.41~$\ms$ and the weighted RV standard
  deviation is 3.54~$\ms$. The ratio between these two values
  clearly suggests that the RVs contain variations not accounted for by the
  uncertainties.

  \subsection{ESPRESSO}

  The new ESPRESSO spectrograph (\citealp{Pepe2014}; 2019 in prep.) started
  operations at the ESO Paranal observatory in September 2018. It is designed and
  built to reach an unprecedented RV precision of 10 $\cms$, with the express
  goal of detecting low-mass rocky planets within the habitable
  zones of their host stars. ESPRESSO is a fibre-fed, cross-dispersed,
  high-resolution échelle spectrograph, combining RV precision and spectroscopic
  fidelity with the large collecting area of the VLT. 

  Within the ESO program with ID
  \href{http://archive.eso.org/wdb/wdb/eso/abstract/query?progid=0102.C-0757}{0102.C-0757}
  (PI: J. Faria), we obtained a total of 23
observations of \target  from November 2018 to March 2019. The measurements were taken in ESPRESSO's
  high resolution 1UT (HR) mode. With the exception of one observation, which was
  obtained with UT1, all other spectra were obtained with UT3. We used a
  Fabry-Perot etalon for simultaneous drift measurement. 
  
  Exposure times were set to 15 minutes to average out the p-mode oscillations
  typical of solar-type stars \citep[e.g.][]{Dumusque2011b,Chaplin2019}.
  According to the ESPRESSO exposure time calculator (version P102), this would
  provide a photon noise precision close to 20 $\cms$. Around 5500~\AA\ (order
  102), the average measured S/N of the ESPRESSO spectra is
  165, while the minimum and maximum values are
  111
and 201.

  As for HARPS, the RVs were extracted from the ESPRESSO spectra with the CCF
  technique, using the publicly available ESPRESSO pipeline (version 1.2.2,
  \citealt{Modigliani2019}). The CCF mask is again optimised for a G2 dwarf, but
  it is different from that of HARPS as it includes a number of additional lines
  in the red part of the spectrum, between 6800 \AA\xspace and 7900 \AA, to
  match ESPRESSO's larger wavelength coverage. 
  
\begin{figure*}
    \resizebox{\hsize}{!}{\includegraphics{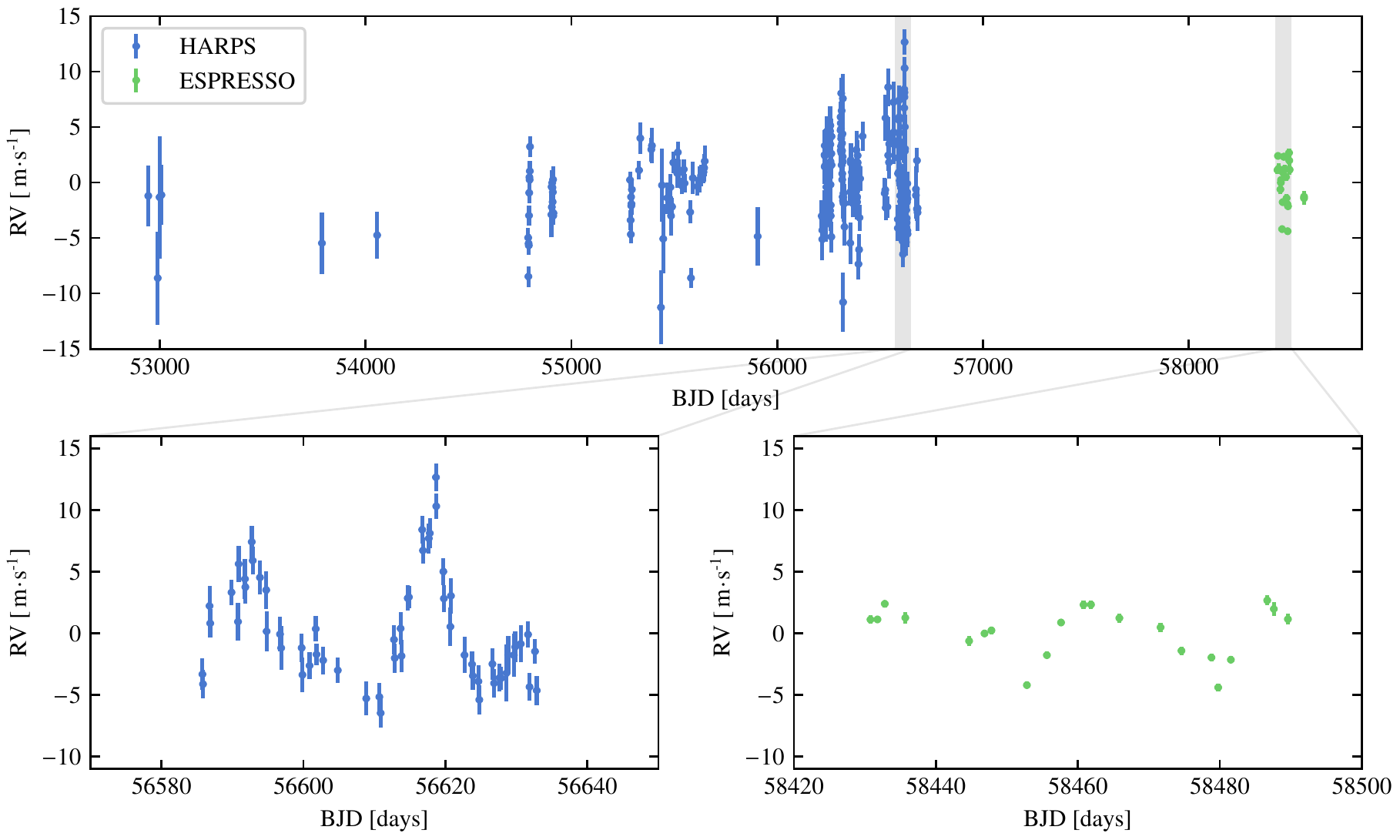}}
    \caption{
      Full set of RVs from HARPS (blue) and ESPRESSO (green).
      Average RV for each data set has been subtracted for visual comparison.
      The two bottom panels show a zoom on a subset of the HARPS observations 
      (left) and the ESPRESSO observations (right). 
      These two panels are shown on the same scale.
      We note that all ESPRESSO observations are shown with error bars
      but in most cases these are smaller than the points.
    }
    \label{fig:RVs}
  \end{figure*}

  Line profile indicators were also calculated for each CCF%
  \footnote{The ESPRESSO pipeline provides values for the FWHM of the CCF. We
  also used the \texttt{iCCF} package, available at
  \href{https://github.com/j-faria/iCCF}{github.com/j-faria/iCCF}, to calculate
  other indicators.}. %
  The average uncertainty on the ESPRESSO RVs is
  36.3
$\cms$, with an rms of
  2.12~$\ms$. We note that the ESPRESSO RV uncertainties only
  include the photon noise contribution. The RV time series is also shown in
  \fig{fig:RVs} (see Table \ref{tab:rvs}) and compared to the HARPS RVs on the
  same scale.

  \subsection{TESS}

  The Transiting Exoplanet Survey Satellite (\TESS; \citealt{Ricker2015}) observes 
   \target (TIC 350844714, \TESS magnitude = 8.187) in
  sectors one through 13 of its nominal two-year mission. As of June 2018, data
  from the first ten
sectors are available (from
  25 July 2018 to 22 April 2019). This leads to a baseline of around
  270
days. \TESS observations are simultaneous with
  the ESPRESSO RVs between the end of sector four and middle of sector nine.

  We downloaded, combined, and analysed the \TESS light curves for the first 10
  sectors. An in-depth analysis of the combined light curve is described in
  Appendix \ref{app:tess}. In summary, we do not detect credible transit signals. We do find evidence for a stellar rotation period between 24 and 25 days. The
  data are consistent with a spot lifetime of about 25 days.

\section{Stellar characterisation}
\label{sec:3}

  \subsection{Fundamental stellar parameters}
  \newcommand{\gaia}{\emph{Gaia}\ }

  \target (\targetothername) is a G2V dwarf in the constellation of Pictor, with
  a visual magnitude of V=8.82. In the \gaia DR2 catalogue
  \citep{GaiaCollaboration2016a,GaiaCollaboration2018a}, its parallax is listed
  as $18.006 \pm 0.028$
mas, leading to a distance of $55.45 \pm 0.09$
parsec, as
  inferred by \citet{BailerJones2018}.

  Stellar atmospheric parameters were derived using both the HARPS and ESPRESSO
  spectra. We use the ARES\,+\,MOOG method for the spectroscopic analysis
  \citep{Sousa2014}. This method is an equivalent width (EW) method, based on
  the ionisation and excitation balance of Fe\,I and Fe\,II lines. The list of
  iron lines is the same as presented in \citet{Sousa2008}. The EWs are
  automatically measured using ARES \citep{Sousa2007} and the iron abundances
  are derived with the MOOG code \citep{Sneden2012}, assuming local
  thermodynamic equilibrium, and with a grid of Kurucz ATLAS9 models
  \citep{kurucz1993}.

  Each HARPS spectrum was shifted by the corresponding RV, after which the flux
  values in each wavelength bin were added. With the combined spectrum, we
  derived the following values for the stellar atmospheric parameters: effective
  temperature T$_{\rm eff}$~=~$5713 \pm 63$~K, surface gravity
  log~g~=~$4.49 \pm 0.10$~dex, and metallicity [Fe/H]~=~$-0.37 \pm 0.05$~dex.
    
  The ESPRESSO spectra were combined in a similar way, finally resulting in
  T$_{\rm eff}$~=~$5724 \pm 13$~K, log~g~=~$4.517 \pm 0.025$~dex,
  and [Fe/H]~=~$-0.348 \pm 0.010$~dex.
  The parameters from HARPS and ESPRESSO agree within the quoted uncertainties.
  Since ESPRESSO provides higher precision -- due to the higher S/N in the
  combined spectra -- we use these values as the final stellar parameters,
  which are listed in Table \ref{tab:parameters}. 
  
  From the stellar atmospheric parameters, we then determined the stellar mass
  and radius using two methods. First, we used the calibrations from
  \citet{Torres2010}. Uncertainties were derived with a Monte Carlo procedure
  which resamples the effective temperature, surface gravity, and metallicity
  using their quoted uncertainties \citep[see][]{Santos2013}. We obtained final
  values of $0.86 \pm 0.06$
\Msun and
  $0.89 \pm 0.04$
\Rsun, respectively.

  As a second method, we derived stellar mass, radius, density, and age with the
  \texttt{isochrones} package \citep{Morton2015} and the Dartmouth stellar
  isochrones \citep{Dotter2008}. As inputs, we used the apparent magnitudes, the
  Gaia parallax, and the spectroscopic parameters. This results in an estimated
  stellar mass of $0.866 \pm 0.015$
\Msun and a radius of
  $0.898 \pm 0.005$
\Rsun, which agree with the
  values from the calibration but are more precise. For the age, we derive a
  value of $6.8 \pm 0.9$
Gyr (see Table \ref{tab:parameters}).

  From a comparison with synthetic spectra, using the code to be described in
  \mbox{Costa Silva et al. (in prep.)}, we also derived the stellar projected
  rotational velocity, $v\sin i$. The method consists in a $\chi^2$ minimisation
  of the deviations between the synthetic profile (created with MOOG) and the
  observed spectra, in several wavelength regions where FeI and FeII lines are
  present. All the stellar parameters are fixed to the values derived
  previously, except for $v\sin i$, which is set as the only free parameter
  affecting the synthetic spectra.
  Our final estimate, using the combined ESPRESSO spectra, is of
  $2.1 \pm 0.8$
$\kms$. Together with the stellar radius, and
  assuming an inclination of 90$^\circ$, this leads to a rotation period of
  $22 \pm 9$
days.

    \begin{table}
      \caption{Stellar parameters and elemental abundances for \target.}
      \label{tab:parameters}
      \centering
      \begin{tabular}{lclc}
      \hline\hline
      \noalign{\smallskip}
      Parameter        & Value & Elem. & Abundance \\
      \hline
      \noalign{\smallskip}
      Spectral~type          & G2V
& C\,I
& $-0.370 \pm 0.024$
\\
      $V$                    & 8.81
& O\,I
& $-0.23 \pm 0.07$
\\
      $B-V$                  & 0.61
& Na\,I
& $-0.311 \pm 0.009$
\\
      $\pi$ [mas]            & $18.006 \pm 0.028$
& Mg\,I
& $-0.320 \pm 0.022$
\\
      Distance~[pc]          & $55.45 \pm 0.09$
& Al\,I
& $-0.310 \pm 0.023$
\\
      M$_V$                  & $5.091 \pm 0.011$
& Si\,I
& $-0.321 \pm 0.031$
\\
      L [L$_{\odot}$]        & $0.7871 \pm 0.0020$
& Ca\,I
& $-0.33 \pm 0.06$
\\
      $\log{R'_{\rm HK}}$    & -4.89
& Sc\,II
& $-0.32 \pm 0.06$
\\
      $T_{\rm eff}$~[K]      & $5724 \pm 13$
& Ti\,I
& $-0.305 \pm 0.031$
\\
      $\log{g}$              & $4.517 \pm 0.025$
& Cr\,I
& $-0.352 \pm 0.028$
\\
      ${\rm [Fe/H]}$         & $-0.348 \pm 0.010$
& Ni\,I
& $-0.391 \pm 0.021$
\\
      $v \sin i$~[$\kms$]    & $2.1 \pm 0.8$
& \\
      Mass~$[M_{\odot}]$     & $0.866 \pm 0.015$
& \\
      Radius~$[R_{\odot}]$   & $0.898 \pm 0.005$
& \\
      Density~$[\rho_{\odot}]$  & $1.197 \pm 0.033$
& \\
      Age~[Gyr]                 & $6.8 \pm 0.9$
& \\
      \hline
      \noalign{\smallskip}
      \end{tabular}
      \end{table}

  We also derived abundances of several elements using the same tools (ARES and
  MOOG) and models (Kurucz ATLAS9) as for the determination of atmospheric
  parameters. For details on the methods, see \citet{Adibekyan2012},
  \citet{Adibekyan2015a}, \citet{DelgadoMena2010}, and \citet{BertrandeLis2015}.
  The derived abundances, measured relative to the Sun, are also listed in Table
  \ref{tab:parameters}.

  \subsection{Activity indices}

  The HARPS and ESPRESSO spectra can be used to calculate several activity
  indices based on the flux in the cores of the Ca\,II H and K, H$\alpha$,
  and Na\,I\,D$_2$ activity-sensitive lines. These indices will be defined
  here as \ica, \iha, and \ina, respectively. 
  
  We used the open-source package \texttt{ACTIN}\footnote{Available at
  \href{https://github.com/gomesdasilva/ACTIN}{github.com/gomesdasilva/ACTIN}.}
  \citep{GomesdaSilva2018} which calculates the activity indices by dividing the
  flux in the core of activity-sensitive spectral lines by the flux in nearby
  reference regions. The code can calculate multiple indices with a given number
  of lines, provided that the line bandpasses fall in the instrument's
  wavelength range. The fractions of pixels inside the band-passes are also taken
  into account (for details, see \texttt{ACTIN}'s documentation).

  An index $I$ is calculated using the expression 
  \begin{eqnarray*}
  I = \sum_{\ell} \sum_{r} \frac{F_\ell}{R_r},
  \end{eqnarray*}
  where $F_\ell$ is the flux in the bandpass centred on the line $\ell$, $R_r$
  is the flux in the bandpass centred at the reference region $r$. The errors of
  the fluxes are considered as Poisson noise, scaling with $\sqrt{N}$, where $N$
  is the photon count in the bandpass. The error of a given index is calculated
  via error propagation and is given by
  \begin{eqnarray*}
  \sigma_I = \frac{\sqrt{ \sum_\ell \sigma_{F_\ell}^2 + I^2 \,  \sum_r \sigma_{R_r}^2}}{\sum_r R_r},
  \end{eqnarray*}
  where $\sigma_{F_\ell}$ and $\sigma_{R_r}$ are the Poisson (photon) errors of
  the fluxes $F_\ell$ and $R_r$, respectively. 
  
  In this work we used the same lines and bandpass parameters as in
  \citet{GomesdaSilva2011}, except for \ica for which we used a triangular
  bandpass with a FWHM of 1.09 \AA\, which simulates the calculation of the Mt
  Wilson $S$-index \citep{Vaughan1978}. Since we are analysing only one
  individual star and a calibration to transform the Ca\,II index to the
  $S$-index scale is still not available for ESPRESSO, we have not computed the
  photospheric-corrected \rhk index%
  \footnote{Note that the HARPS pipeline does provide the \rhk index.} %
  \citep{Noyes1984}.

\begin{figure*}
    \includegraphics[width=\hsize]{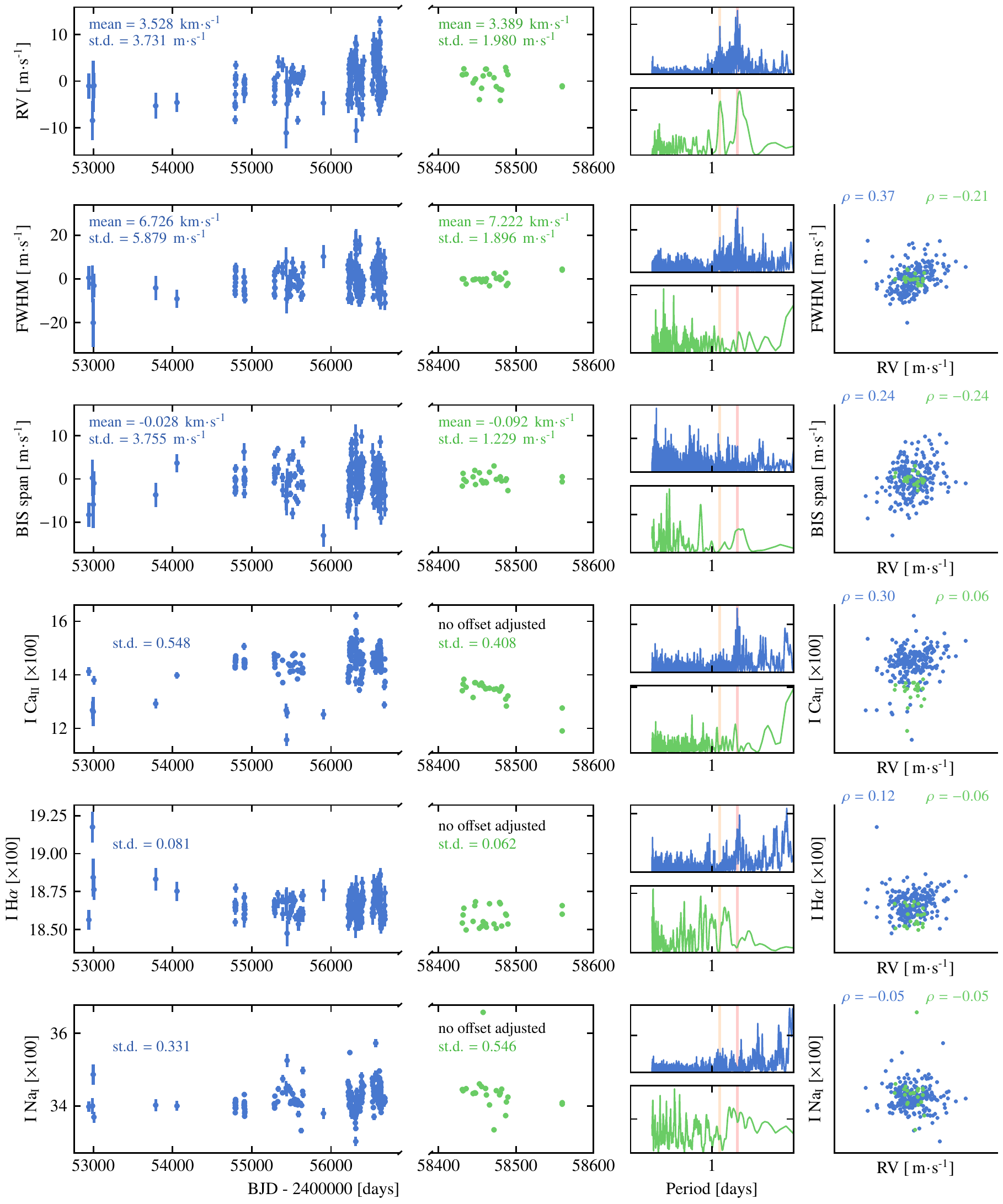}
    \caption{\textit{Left:} Time series of RVs, CCF FWHM, CCF bisector span,
    \ica, \iha, and \ina for HARPS (blue) and ESPRESSO (green). The vertical
    scales are the same in both broken panels. For the RVs, FWHM, and BIS span
    an offset between the two instruments was adjusted and subtracted. The
    average values before this offset are shown in the panels. \textit{Middle:}
    Bayesian Lomb-Scargle periodograms of the HARPS and ESPRESSO observations.
    The red lines indicate the periods of 13.3 and 25.6 days. \textit{Right:}
    Correlation of each activity indicator with the RVs, showing Pearson's
    correlation coefficient individually for the two instruments.}
    \label{fig:activity_all}
  \end{figure*}

  Figure \ref{fig:activity_all} shows the time series of RVs, FWHM, bisector
  span, \ica, \iha, and \ina for ESPRESSO and HARPS. In the case of the RVs,
  FWHM, and bisector span, an offset between the two instruments was adjusted,
  since the absolute value of these observables should have an instrumental
  dependence. The average values of each quantity before adjusting this offset
  are also shown in the panels and presented in Table \ref{tab:offsets}.
  
    \begin{table}
    \caption{Offsets between HARPS and ESPRESSO.}
    \label{tab:offsets}
    \centering
    \begin{tabular}{lccc}
    \hline \hline
    Quantity & HARPS & ESPRESSO & offset            \\
             & mean   & mean    & HARPS - ESPRESSO  \\
             & $[\kms]$ & $[\kms]$  & $[\ms]$       \\
    \hline
    RV       & 3.52808
& 3.38867
& 139.41
\\
    FWHM     & 6.7258
& 7.22179
& -495.99
\\
    \hline
    \end{tabular}
    \end{table}

  The Bayesian Generalised Lomb-Scargle (BGLS) periodograms \citep{Mortier2015}
  of the different quantities are shown in the middle panels, individually for
  HARPS and ESPRESSO, and the right panels show the correlation between the RVs
  and the other variables, with an indication of Pearson's coefficient of
  correlation.

\section{Analysis of the radial velocities}
\label{sec:4}

  To analyse the two RV data sets, we will consider, in turn, white and
  correlated noise models. Details on the models and the inference procedure are
  given below.

  \subsection{RV model}

  As a common component of both white and correlated noise models, we assume that
  the RV variations caused by planets can be modelled as a sum of Keplerian
  curves. Each Keplerian can be described with five parameters: the
  semi-amplitude $K$, the orbital period $P$, the eccentricity $e$, the time of
  periastron%
  \footnote{Parameterised here with the mean anomaly, $M$.} %
  $T_p$, and the argument of periastron, $\omega$. A model with $N_p$
  planets corresponds to the sum of the $N_p$ individual Keplerian curves.

  The RV zero-point corresponds to the radial velocity of the centre of mass of
  the system, here denoted by $v_{\rm sys}$. Since the RVs come from two
  instruments, we need to account for an offset, denoted $\delta$ here, between
  the HARPS and ESPRESSO measurements%
  \footnote{In practice, we consider HARPS as the reference data set (since it
  has more points) including an offset for the ESPRESSO data (cf. Eq.
  \ref{eq:RV_model}).}.
  In order to model any long-term trends present in the RVs, we also include a
  linear term (with slope $\beta$ and referenced to $t_{\rm ref}$, the average
  time of the observations) which is shared between the two instruments.
  
  In summary, for a model with $N_p$ planets, the model for the RV of the star
  at time $t$ is calculated as
  
  \begin{align}
    v (t) = 
          v_{\rm sys}
          &+ \delta \,{\rm I}_{\,t\,\in\,\rm ESP}
          + \beta (t-t_{\rm ref}) \notag \\
          &+ \sum\limits_{\scriptscriptstyle i=1}^{\scriptscriptstyle N_p} %
            K_i \left\{ \cos\left[\omega_i + f_i(t)\right] + e_i \cos \omega_i \right\} %
  \label{eq:RV_model}
  \end{align}

  where ${\rm I}_{\,t\,\in\,\rm ESP}$ is an indicator variable equal to 1 if
  time $t$ belongs to the ESPRESSO data set and 0 otherwise, and $f_i(t)$ is the
  true anomaly for planet $i$, which can be calculated from the other orbital
  parameters \citep[see e.g.][]{Perryman2014}. 

  In addition to the Keplerian signals, the measured RVs can include variations
  originating from other sources, namely stellar activity or instrumental noise.
  Here we formulate three different models for these sources of noise by
  defining the appropriate likelihood function or adding an additional term to
  Eq. \ref{eq:RV_model}.

  As a simple white noise model,
  for a data set $\mathcal{D}$ containing times $t_k$, 
  radial velocities $v^{\rm obs}_k$, and uncertainties $\sigma_k$,
  we use an independent Gaussian likelihood
  \begin{equation}
    \label{eq:likelihood}
    \mathcal{L}(\theta) = \prod_k \frac{1}{\sqrt{2\pi \left(\sigma_k^2 + s^2\right)}}
                                  \exp \left[ - 
                                    \frac{\left(v^{\rm obs}_k - v(t_k)\right)^2}%
                                         {2 \left(\sigma_k^2 + s^2\right)} 
                                       \right] ,
  \end{equation}
  where $\theta$ is a vector containing all the unknown parameters
  $\left\{v_{\rm sys}, \delta, \beta, P, K, e, \omega, T_p, 
  s\right\}$. The additional white noise component controlled by the parameter
  $s$ (often called jitter) can accommodate intrinsic stellar variability or
  instrumental effects which are not included in the RV uncertainties. In
  practice, we consider individual jitter parameters for HARPS and for ESPRESSO.

  A further generalisation is to allow for off-diagonal terms in the Gaussian
  likelihood, effectively modelling the RVs with a Gaussian process (GP). The
  log-likelihood is then given by 
  \begin{align}
    \ln \mathcal{L}(\theta) = -\frac{1}{2} \mathbf{r}^T \, \Sigma^{-1} \, \mathbf{r} 
                                - \frac{1}{2} \ln \det \Sigma
                                - \frac{N}{2} \ln 2\pi \text{,}
    \label{eq:likelihood-gp}
  \end{align}
  where $\mathbf{r}$ is the vector of residuals, given by $v^{\rm obs}_k -
  v(t_k)$, and $\Sigma$ is the covariance matrix. This matrix is obtained by
  evaluating the covariance function (or kernel) of the GP at the observed
  times.
  From the many possible choices for a covariance function, 
  the quasi-periodic kernel is the most widely used in the exoplanet literature 
  \citep[e.g.][]{Haywood2014,Faria2016,Cloutier2017},
  resulting in a covariance matrix of the form:
  \begin{equation}\label{eq:qp-kernel}
    \Sigma_{ij} = \eta_1^2 
        \exp\left[ - \frac{(t_i - t_j)^2}{2\eta_2^2} 
                  - \frac{2\sin^2\left(\frac{\pi (t_i-t_j)}{\eta_3}\right)}{\eta_4^2} \right] 
        + \left( \sigma_i^2 + s^2 \right) \delta_{ij} \:\text{.}
  \end{equation}

  \noindent
  In this GP model, $\eta_1$, $\eta_2$,  $\eta_3$, and $\eta_4$ are free
  hyperparameters. They correspond, respectively, to the amplitude, timescale of
  decay, periodic timescale, and level of high-frequency variability in the GP
  model \citep[see][]{Rasmussen2006}. Both $\eta_2$ and $\eta_3$ may have
  physical interpretations as the timescale for evolution of active regions in
  the stellar surface, and the stellar rotation period. 
  The parameter $\eta_4$ controls the scale of variability of the GP within one
  period $\eta_3$, with larger values corresponding to more sinusoidal
  functions.
  As before, there is one scalar jitter parameter for each instrument,
  while the remaining parameters are shared.

  A third possibility is to consider a first-order moving average (MA) term in
  the RV model, as done by \citet{Jenkins2014} for \target and several other
  authors \citep[e.g.][]{Tuomi2014,Diaz2018a,Jenkins2017,Feng2016}. In the MA
  model, we consider the following additional term added to the right-hand-side
  of Eq. \eqref{eq:RV_model}:

  \begin{align}
    v (t_k) = \ldots 
            + \phi \, \exp\left( \frac{t_{k-1} - t_k}{\tau} \right) \, r_{k-1} ,
  \label{eq:MA}
  \end{align}

  where $r_{k-1}$ denotes the residuals after subtracting the model from the
  $(k-1)$th measurement%
  \footnote{The MA component of the RV model in Eq.~\eqref{eq:MA} is only
  defined at the observed times $t_k$, unlike the other terms from
  Eq.~\eqref{eq:RV_model}.}.

  This term can account for correlations between consecutive observations, with
  exponential smoothing over a timescale of $\tau$ days. The amplitude of the
  correlations is set by the parameter $\phi$, with values between -1 and 1.
  Both $\phi$ and $\tau$ are free parameters, shared between HARPS and ESPRESSO.

  \subsection{Prior distributions}

    We use informative but broad prior distributions for the model parameters.
    The orbital periods are assigned a log-uniform (often called Jeffreys) prior
    between one day and twice the time span of the full data set ($\Delta t \sim
    5615$ days). We consider modified log-uniform
    distributions for both the semi-amplitudes and the jitter parameters $s$,
    limited above by the variance%
    \footnote{The variance was calculated after subtracting the mean RV from
    each instrument and with the degrees of freedom set to 1.} %
    of the RV observations (${\rm var} \: v^{\rm obs} =13.05
\ms$)
    and with an inflection at 1$\ms$. These modified distributions are defined
    until the lower limit of $0 \ms$, which explicitly allows for the Keplerian
    and additional white noise terms to cancel.
    
    It is also worth nothing the prior for the orbital eccentricities, which is the
    Kumaraswamy distribution \citep{Kumaraswamy1980}, with shape parameters
    $\alpha=0.867$ and $\beta=3.03$. This distribution closely matches the Beta
    distribution proposed by \citet{Kipping2013}, which is used as an
    approximation to the frequency distribution of exoplanet eccentricities. The
    Kumaraswamy distribution has a small computational advantage since its
    cumulative distribution function can be easily evaluated. Other parameters
    are assigned uniform priors between sensible limits.
    
    For the GP hyperparameters, we use relatively broad priors: uniform
    distributions in the logarithm of $\eta_1$ and $\eta_4$, a log-uniform
    distribution for $\eta_2$ and uniform for $\eta_3$ between appropriate
    limits for a solar-like star (see Table \ref{tab:priors}). The most
    restrictive of these priors is the one for $\eta_3$, associated with the
    rotation period of \target, which we define between ten and 40 days. In any
    case, all evidence gathered so far points to these limits being appropriate.
    We note that the prior for $\eta_1$ can be considered too wide given the
    observed RV dispersion. However, its functional form assigns low probability
    to large values of the parameter, and we find that our results do not depend
    strongly on the prior upper limit. The MA parameters $\phi$ and $\tau$ are
    also assigned broad but meaningful priors. In particular, we allow for
    $\tau$ to be as large as 100 days to model activity-related signals.

    \begin{table}
    \caption{Prior distributions for the parameters in the RV model.}
    \label{tab:priors}
    \centering    
    \begin{tabular}{lcr}
    \hline\hline\noalign{\smallskip}
    Parameter & Unit & Prior distribution \\
    \hline\noalign{\smallskip}
    $N_p$     &   & $\mathcal{U}\,(0, 4)$                           \\ [1ex]
    $P$        & days   & $\mathcal{LU}\,(1, 2 \Delta t)$      \\
    $K$        & $\ms$  & $\mathcal{MLU}\,(1, {\rm var} \: v^{\rm obs} )$       \\
    $e$        &        & $\mathcal{K}\,(0.867, 3.03)$              \\
    $M \left[=\frac{2\pi}{P}(t_0 - T_{\!p})\right]$     &        & $\mathcal{U}\,(0, 2\pi)$                  \\
    $\omega$   &        & $\mathcal{U}\,(0, 2\pi)$                  \\[1ex]
    %
    $\ln \eta_1$    & $\ms$   & $\mathcal{U}\,(-5, 5)$       \\
    $\eta_2$        & days    & $\mathcal{LU}\,(1, 100)$     \\
    $\eta_3$        & days    & $\mathcal{U}\,(10, 40)$      \\
    $\ln \eta_4$    &         & $\mathcal{U}\,(-1, 1)$       \\[1ex]
    $\phi$          & $\ms$   & $\mathcal{U}\,(-1, 1)$       \\
    $\tau$          & days    & $\mathcal{LU}\,(1, 100)$     \\[1ex]
    $s$           & $\ms$  & $\mathcal{MLU}\,\left(1, V \right)$                  \\
    $v_{\rm sys}$ & $\ms$  & $\mathcal{U}\,(\min v^{\rm obs}, \max v^{\rm obs})$  \\
    $\beta$       & $\msy$  & $\mathcal{U}\,(- \frac{\Delta v^{\rm obs}}{\Delta t}, \frac{\Delta v^{\rm obs}}{\Delta t})$  \\[.2em]
    $\delta$      & $\ms$  & $\mathcal{U}\,(- \Delta v^{\rm obs}, \Delta v^{\rm obs})$                               \\
    \end{tabular}
    \tablefoot{%
               $\Delta t$ and ${\rm var} \: v^{\rm obs}$ are the time span and variance of the RVs, respectively,
               while $\Delta v^{\rm obs}$ is the total span of the RVs.
               $\mathcal{U}(\cdot, \cdot)$ is a uniform prior (discrete in the case of \np) with lower and upper limits;
               $\mathcal{LU}(\cdot, \cdot)$ is a Jeffreys (log-uniform) prior with lower and upper limits;
               $\mathcal{MLU}(\cdot, \cdot)$ is a modified Jeffreys prior with knee and upper limit;
               $\mathcal{K}(\alpha, \beta)$ is a Kumaraswamy prior with shape parameters $\alpha$ and $\beta$.
               See, e.g. \citet{Ford2006,Gregory2005}. 
               }
    \end{table}

  \subsection{Inference}

    For the analysis of Section \ref{sec:pre-whitening}, we will simply maximise
    the log-likelihood function instead of sampling from the posterior
    distribution. This was done by combining several of the optimisation methods
    implemented in the \texttt{SciPy} package \citep{Jones2001}. The GLS
    periodograms were calculated using the \texttt{astropy} package
    \citep{astropy:2018}.

    To sample from the posterior distribution for the model parameters, we use
    the diffusive nested sampling (DNS) algorithm proposed by
    \citet{Brewer2011}. In addition to posterior samples, DNS also provides an
    estimate of the Bayesian evidence, $Z$, the constant which normalises the
    posterior distribution.
    The value of $Z$ can be used for model comparison
    \citep[e.g.][]{Gregory2011,Feroz2011}. In our case, we may want to compare
    models with different numbers of planets. For example, for two models with
    $N_p=0$ and $N_p=1$, with equal prior probabilities, the logarithm of the
    Bayes factor is given by
    \begin{equation}
      \ln \mathcal{B}_{1,0} = \ln Z_1 - \ln Z_0
    \end{equation}
    where $Z_0$ is the evidence of the first model and $Z_1$ that of the second.
    The Bayes factor $\mathcal{B}$ is a continuous measure of evidence for (or
    against) one model relative to another. It can be interpreted with the scale
    introduced by \citet{Jeffreys1961} -- see also \citet{Feroz2011} and
    \citet{Efron2001}. In short, we require Bayes factors of at least 150 ($\ln
    \mathcal{B} \simeq 5$) between consecutive models in order to claim a planet
    detection \citep{Faria2016}. A more conservative criterion would be to
    require the Bayes factors to reach 1000 ($\ln \mathcal{B} \simeq 6.9$).

    All analyses were carried out with \texttt{kima}, an open-source package%
    \footnote{Available at
    \href{https://github.com/j-faria/kima}{github.com/j-faria/kima}.}\,%
    aimed at the analysis of RV data sets \citep{Faria2018}. The code can
    calculate the evidence for a model with a given number of planets or
    estimate the posterior distribution for $N_p$. In both cases, the posteriors
    for the remaining parameters are also estimated in one single run. This
    includes the parameters of the GP or the MA components.

\section{Results}
\label{sec:5}

  \subsection{Pre-whitening procedure}
  \label{sec:pre-whitening}

    A common procedure for a first analysis of RV data sets is to fit and
    subtract models with a successive number of Keplerians, assessing
    statistical significance with the periodogram false-alarm probability (FAP).
    Using the combined HARPS + ESPRESSO data set, we first consider a white noise
    model with $\np=0$ and maximise the log-likelihood, as given by Eq.
    \eqref{eq:likelihood}. The optimised model is then subtracted from the RV
    data, and the GLS periodogram is calculated for the residuals. The FAP 10\%,
    1\%, and 0.1\% levels are determined by a bootstrap procedure.

    In the periodogram of the residuals, the highest peak is at 25.63 days (with
    FAP $\ll 0.1\%$), surrounded by a forest of peaks. A second region with
    periods between 10 and 20 days is dominated by a peak at 13.36 days (FAP
    $\ll 0.1\%$). Some power is also seen between 8 and 9 days, but these peaks
    are below the 10\% FAP level. We note these periods are close to being
    harmonics of each other \citep[see, e.g.][]{Boisse2011}.

    We then repeat the optimisation of the log-likelihood function with a model
    having $\np=1$. As initial guesses we use the parameters found for the
    previous model and extract from the periodogram initial guesses for the
    Keplerian parameters. The resulting maximum-likelihood values for some of
    the Keplerian parameters are $P=25.61$ days, $K=3.05 \ms$, $e=0.41$. The
    optimised model is then subtracted from the RV data.
    The periodogram of the residuals does not show any peaks above 1\% FAP, with
    the highest one being at 21.84 days.

    Despite the lack of significant peaks, attempting a fit with $\np=2$ results
    in a solution with orbital periods at $P_1 = 25.62$ days and $P_2 = 13.35$
    days. The semi-amplitudes converge to $K_1=2.47\ms$ and $K_2=1.56 \ms$, with
    a much lower eccentricity for the first signal ($e_1=0.08$) and a significant eccentricity for the second ($e_2=0.23$).

  \subsection{Search for Keplerian signals}

  We continue the analysis of the RVs with a blind search for Keplerian signals
  on the combined data set using the white noise model. First, we run models
  with the number of planets \np fixed to 0, 1, 2, 3, and 4, successively. The
  priors for all parameters are the same in all runs, as listed in Table
  \ref{tab:priors}. The estimated evidences for each value of \np are shown in
  Table \ref{tab:evidences}, together with the Bayes factors between models with
  consecutive values of $N_p$.

  \begin{table}[h!] 
    \caption{Logarithm of the evidence and Bayes factors between white noise 
    models for successive values of $N_p$.}
    \label{tab:evidences}
    \centering
    \begin{tabular}{l c c}
    \hline\hline
    \noalign{\smallskip}
    \np & $\ln Z$ & $\ln \mathcal{B}_{i+1, i}$ \\
    \hline
      0 & $-680.63$ & \\
        & & $44.0$ \\
      1 & $-636.63$ & \\
        & & $6.44$ \\
      2 & $-630.18$ & \\
        & & $5.1$ \\
      3 & $-625.09$ & \\
        & & $2.28$ \\
      4 & $-622.81$ & \\
    \hline
    \noalign{\smallskip}
    \end{tabular}
    \end{table}
  
  The highest evidence value corresponds to the $\np=4$ model, but the Bayes
  factors between consecutive models are larger than the detection threshold
  only up to $\np=3$ ($B_{3,2}$). Therefore, this analysis leads to the
  significant detection of three signals.

\begin{figure}
    \centering
    \includegraphics{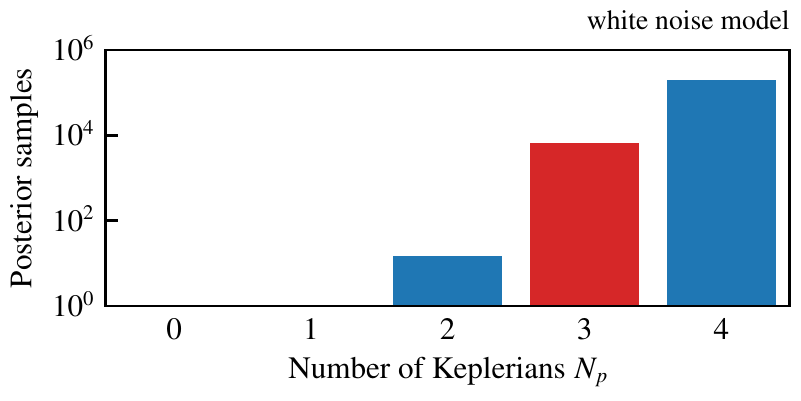}\\
    \includegraphics{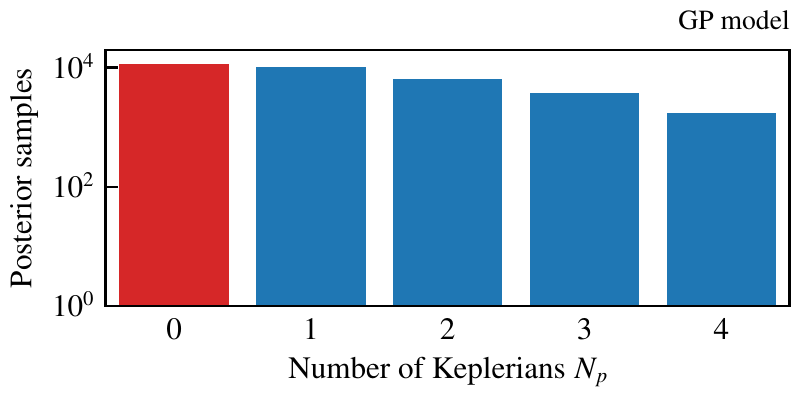}\\
    \includegraphics{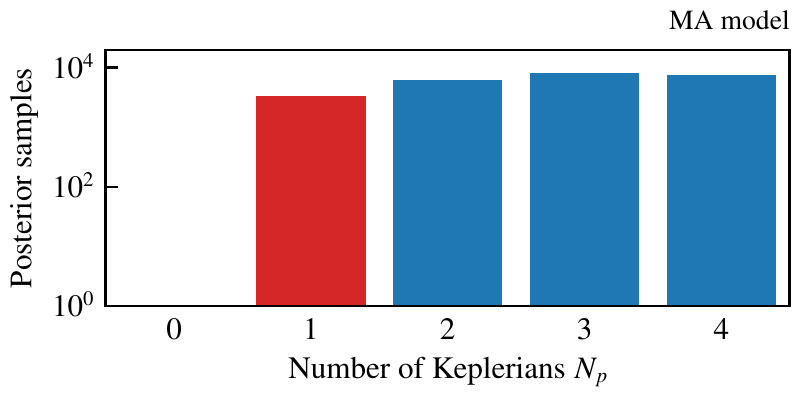}
    \caption{Posterior distributions for \np from the white noise model (top),
      the GP model (middle), and the MA model (bottom). The distributions are
      shown as the number of posterior samples for each value of \np (in
      logarithmic scale). The red bar highlights the largest value of \np for
      which the $\mathcal{B}_{i+1,\,i} > 150$, corresponding to a significant
      detection.}
    \label{fig:np_posterior}
  \end{figure}

  We repeat a similar analysis but with \np free within the uniform prior given
  in Table \ref{tab:priors}. Its posterior distribution is shown in the top
  panel of \fig{fig:np_posterior} and leads to the same conclusion: the
  significant detection of three Keplerian signals. The final evidence for this
  model is estimated as $\ln Z=-622.35$.
  
\begin{figure}
    \includegraphics[width=\hsize]{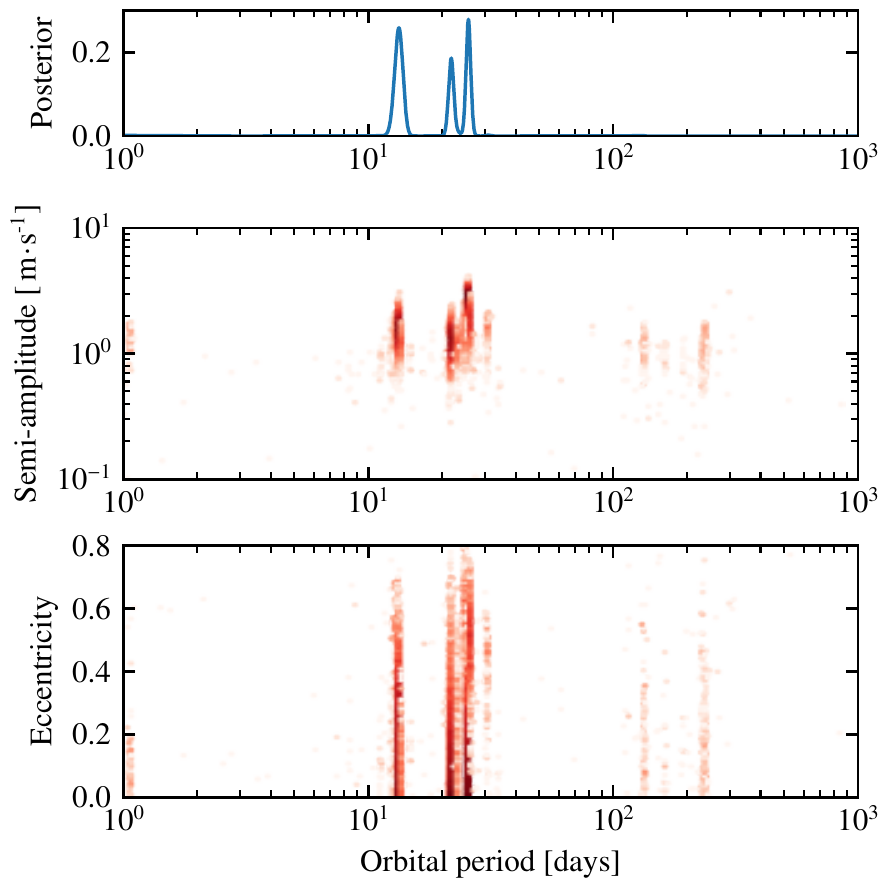}
    \caption{Posterior distributions for the semi-amplitudes, eccentricities and
      orbital periods within the $\np=3$ model with white noise.}
    \label{fig:k3_2d_posterior}
  \end{figure}

  For simplicity, we will now consider only the results from the model with \np
  fixed to 3. The posterior distributions for the orbital periods,
  semi-amplitudes and eccentricities are shown in \fig{fig:k3_2d_posterior}. The
  three highest peaks in the posterior have periods close to 13.3, 21.8 and 25.6
  days and amplitudes around 1.5, 1.2, and 2.7 $\ms$, respectively.

\begin{figure*}
    \includegraphics[width=\hsize]{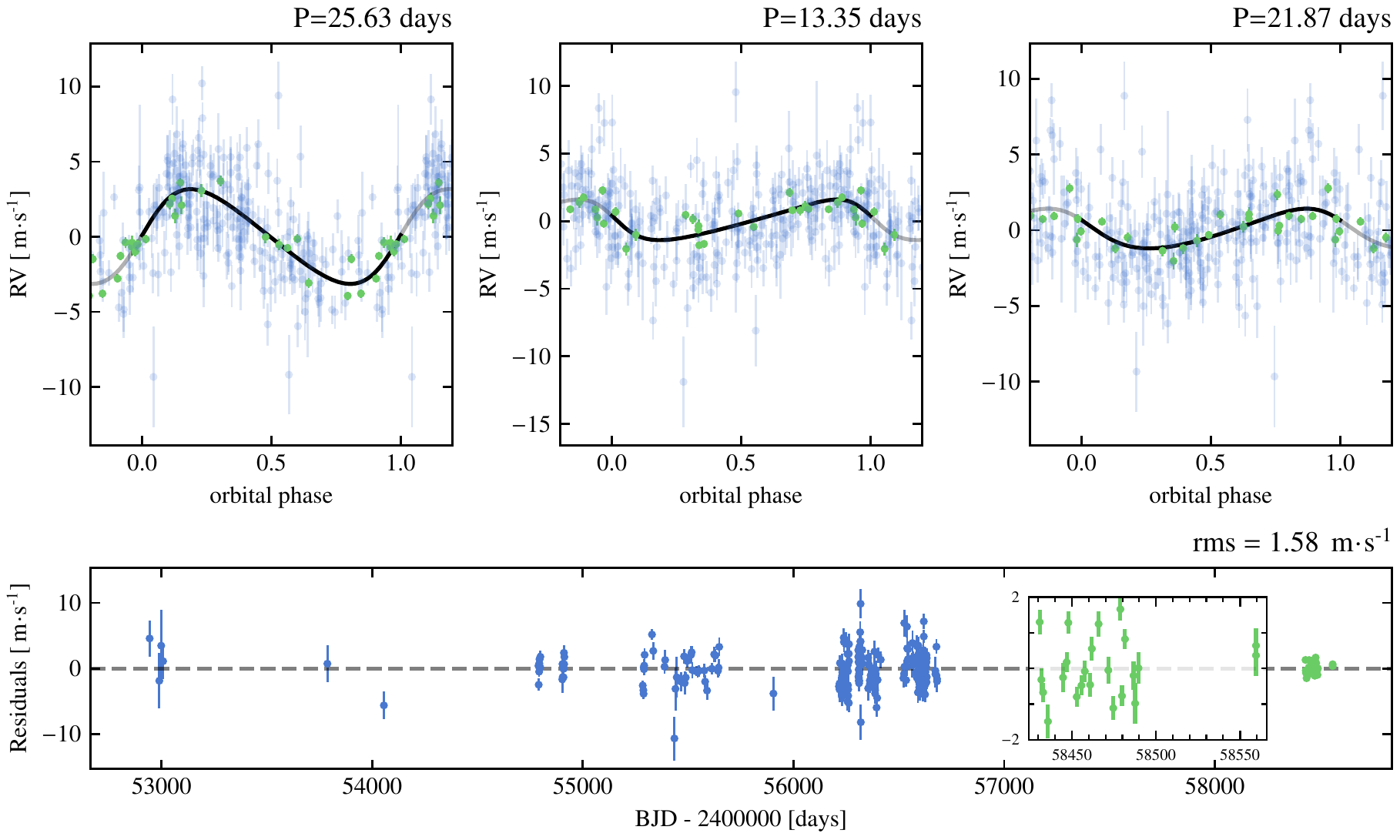}
    \caption{Phase plots and residuals for the maximum likelihood solution with
    $\np=3$ and a white noise model. The top panels show the RVs folded at each
    of the three orbital periods, after removing the contribution from the other
    two Keplerians. The bottom panel shows the residuals after subtracting the
    combined signal of the three Keplerians, with an inset showing the ESPRESSO
    residuals. Data from HARPS and ESPRESSO are colour coded as in \fig{fig:RVs},
    but in the top panels the ESPRESSO points are highlighted. The residual
    (weighted) rms for the full data set is also shown.}
    \label{fig:phasek3}
  \end{figure*}

  Based on the maximum likelihood solution with $\np=3$, the phase folded
  Keplerian curves are shown in \fig{fig:phasek3}, together with the residuals
  for the full data set. The rms of the residuals is of 1.58
$\ms$, which can be compared to the average RV uncertainty of
  1.31
$\ms$. Individually for the two
  instruments, the HARPS residuals show an rms of
  2.17
$\ms$
  and the ESPRESSO residuals of 0.9
$\ms$. These values are in agreement with the
  posterior estimated for the jitter parameters: $2.01 ^{+0.16} _{-0.15}$$\ms$ and $0.86 ^{+0.23} _{-0.18}$
$\ms$,
  for HARPS and ESPRESSO respectively.

  If these three Keplerian signals are caused by planets, their minimum masses
  can be estimated using this maximum likelihood solution and the value for the
  stellar mass derived above. This results in 4.5,
  12.2, and 5.7
\Mearth, for the three
  periods respectively.

  \subsection{Including a GP model for stellar activity}
  \label{subsec:GPmodel}

    We now attempt to model both RV data sets by including quasi-periodic
    correlated noise. We use the GP model described by Eq.
    \ref{eq:likelihood-gp} with the kernel from Eq. \ref{eq:qp-kernel}. The
    number of planets is free and the prior distributions are those of Table
    \ref{tab:priors}.
    After obtaining 100\,000 samples from DNS's target distribution, this
    corresponds to 34264
posterior samples. The posterior distribution for
    \np is shown in the middle panel of \fig{fig:np_posterior} and does not
    result in a significant detection of any Keplerian signals. The final
    evidence for this model is estimated as $\ln
    Z=-606.32$.
    
\begin{figure}
      \includegraphics[width=\hsize]{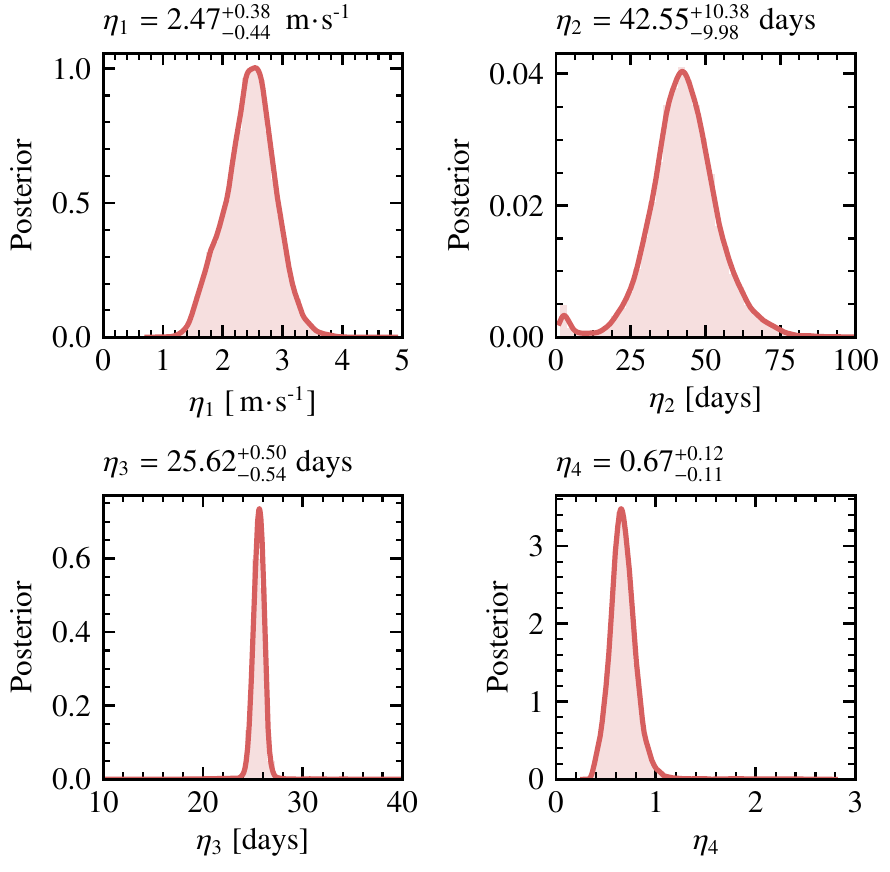}
      \caption{Posterior probability densities for the GP hyperparameters within
      the model where \np is free. These distributions combine samples from the
      posterior for all values of \np. Each panel also indicates the median and
      68\% quartiles of the distribution.}
      \label{fig:posteriors_etas}
    \end{figure}

    The GP alone is able to reproduce the RV variations of the full data set.
    The resulting posterior distributions for the hyperparameters of the GP are
    shown in \fig{fig:posteriors_etas}. For $\eta_3$, the posterior is clearly
    peaked at 25.62 days, which may correspond to the stellar rotation period.
    The amplitude of the correlations ($\eta_1$) is estimated around
    $2.45\ms$, close to the semi-amplitude of the 25-day
    Keplerian signal found before. Moreover, $\eta_2$ is relatively well
    constrained around $42.5$ days, which can be
    physically interpreted as the timescale for active region evolution
    \citep[e.g.][]{Giles2017}. Finally, the posterior estimate of
    0.67
for $\eta_4$ means that the RV curve, as
    modelled by the GP, shows variability on a timescale of about half the
    stellar rotation period. This can be explained as the active regions on the
    stellar surface go in and out of view as the star rotates
    \citep[e.g.][]{Lopez-Morales2016,Haywood2016a}.

\begin{figure}
      \centering
      \includegraphics[width=\hsize]{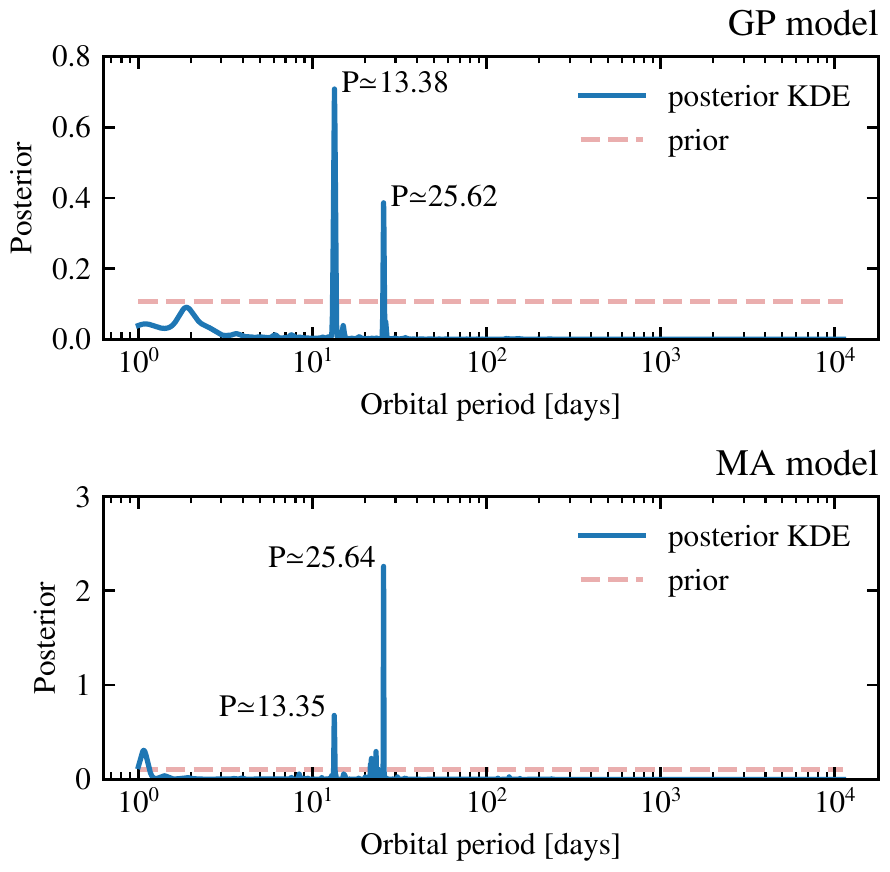}
      \caption{Posterior for the orbital periods from the GP and MA models. A
        kernel density estimation (KDE) of the posterior samples is shown
        together with the probability density function of the prior. We note that
        the height of the peaks relative to the prior is not a measure of
        significance.}
      \label{fig:GP_MA_period_posteriors}
    \end{figure}

    Even if they are not significant enough to provide a detection, the samples with $\np
    \ge 1$ still have substantial posterior probability within the GP model. The
    highest peak in the combined posterior for the orbital periods, shown in the
    top panel of \fig{fig:GP_MA_period_posteriors}, is around 13.38 days,
    followed by a peak around 25.62 days. No other period clearly stands out in
    the posterior distribution. It is important to note that in the samples
    where $\np \ge 1$, the GP coexists with the Keplerian curves at these two
    orbital periods. No clear systematic difference is seen in the posteriors
    for the GP hyperparameters as a function of \np, although $\eta_1$ decreases
    slightly with an increasing number of Keplerians.

  \subsection{Moving average model}

      In this Section, we consider a second model for correlated noise which
      uses the MA term from Eq. \eqref{eq:MA} together with the likelihood from
      Eq. \eqref{eq:likelihood} (and, therefore, no GP). Again, the complete
      data set is analysed, with \np free within its uniform prior. The MA
      parameters $\phi$ and $\tau$ are assigned the priors from Table
      \ref{tab:priors} and estimated together with the remaining parameters.
      These two parameters are the same for HARPS and ESPRESSO.
      
      We obtained 100\,000 samples from DNS's target distribution, resulting in
      25689
posterior samples. The final posterior distribution for \np is
      shown in the bottom panel of \fig{fig:np_posterior}, providing evidence
      for the detection of only one Keplerian signal, according to our detection
      criterion. Figure \ref{fig:GP_MA_period_posteriors} (bottom panel) shows
      the posterior for the orbital periods within the MA model, where the
      highest peak is again close to 25.6 days, followed by a less significant
      peak at 13.3 days. 
      
      Choosing all samples having $\np=1$ provides the following posterior
      estimates: orbital period $P=$
      $25.61 ^{+0.01} _{-0.01}$
days, semi-amplitude $K=$
      $2.86 ^{+0.22} _{-0.27}$
$\ms$, and eccentricity $e=$
      $0.35 ^{+0.07} _{-0.09}$.
      The two MA parameters $\tau$ and $\phi$ do not necessarily have physical
      interpretations. From the full posterior distribution, $\phi$ is
      relatively well constrained as
      $0.27 ^{+0.17} _{-0.28}$~$\ms$, but $\tau$ is
      mostly unconstrained within the prior.

\section{Discussion}
\label{sec:6}

    In this Section, we discuss the main implications of our results and attempt to provide
    a unified explanation for the observed RV variations.

    \subsection{Clues from the data}

      \begin{figure}
        \centering
        \includegraphics[width=\hsize]{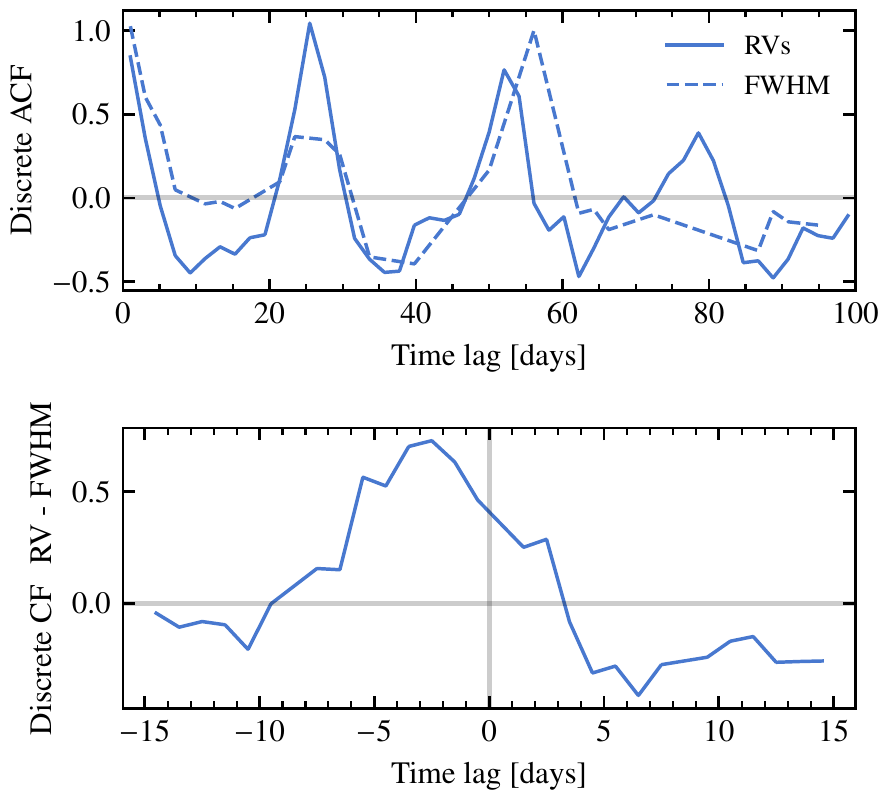}
        \caption{Discrete auto-correlation and cross-correlation functions (top
        and bottom panels, respectively) of the RVs and the FWHM of the CCF for
        the HARPS data set.}
        \label{fig:acf_harps}
      \end{figure}

      Going back to the beginning, the bottom panels of \fig{fig:RVs} show parts
      of the HARPS and ESPRESSO data sets on a common scale, both in RV and time.
      The difference in RV precision between the two instruments is clear. What is also
      evident is that the ESPRESSO RVs show a lower amplitude variation: the HARPS
      peak-to-peak (PTP) RV variations during this time amount to
      19.13
$\ms$, while for ESPRESSO the PTP RV
      variation is only 7.07
$\ms$. It is
      unlikely that one single Keplerian signal could cause these variations, even
      considering instrumental noise.
    
      The BGLS periodograms in \fig{fig:activity_all} (middle panels) show that
      both the HARPS and ESPRESSO RVs contain periodicities around 25 and 13 days.
      We note, however, that the main peak in the periodogram of the ESPRESSO RVs
      is closer to 28 days, and quite wide. At face value, this could be seen as
      evidence for one or two planetary companions. However, the HARPS data also
      show telltale signs of stellar activity: the FWHM of the CCF, the \ica
      index, and the \iha index all share the same periodicity close to 25 days.
      
      The ESPRESSO activity indicators do not show the same clear periodicity and
      also show smaller dispersion, when compared with HARPS. This can be
      explained if \target is now at a lower activity level of a long-term
      magnetic cycle, as suggested by the smaller values and general trend of the
      \ica index. If this is the case, and the RV variations are caused by
      activity, the smaller RV dispersion that we observe with ESPRESSO is
      expected \citep[see, e.g.][]{Diaz2016a}. Part of the additional dispersion
      on some of the HARPS indicators (e.g. FWHM and BIS) could also be due to
      photon noise. However, the ratio (HARPS over ESPRESSO) between the standard
      deviations of the indicators ($\sim$ 3) is higher than the ratio of average
      S/N in the spectra ($\sim$ 2). Also, the FWHM of the HARPS CCF clearly shows
      activity-related variations during part of the observations (see
      \citealt{Santos2014}). We then attribute the observed smaller dispersion
      mostly to a quieter stellar activity phase.
    
      Additionally, the FWHM of the HARPS CCF presents a time lag relative to the
      RVs, already identified by \citet{Santos2014}. We represent this time lag by
      calculating the discrete auto- and cross-correlation functions between the
      RVs and the FWHM, using the algorithm from \citet{Edelson1988}. The results
      are shown in \fig{fig:acf_harps}. The auto-correlation functions show that
      both RV and FWHM share a similar periodicity around 25 days, as is already
      evident from the periodograms. The cross-correlation function reveals that
      the maxima in RV occurs $\sim$3 days before the maxima in FWHM. A very
      similar behaviour has been recently identified for the Sun
      \citep{CollierCameron2019} and is directly connected to the presence of active
      regions on the solar surface.

      The existence of time lags between the observed RVs and some activity
      indicators weakens their mutual linear correlations, which explains the
      small observed values of the correlation coefficients (see
      \fig{fig:activity_all}, right panels). It may also explain why the model
      employed by \citet{Jenkins2014} and \citet{Feng2017b}, which includes these
      linear correlations, may not be effective at correcting for activity-induced
      RV signals.

    \subsection{Results from different noise models}

      We have analysed the joint RV data set from HARPS and ESPRESSO considering
      three noise models of different complexity and, for each of them, we estimated
      how many Keplerian signals are significantly detected in the RV data. The
      outcome differs from one model to the other: the white noise model, GP
      model, and MA model lead to the detection of 3, 0, and 1 Keplerian signals,
      respectively (see \fig{fig:np_posterior}). Given these results, we must now
      compare the models so that we can assess which one (if any) provides a
      better description of the available data.

      \begin{figure}
        \centering
        \includegraphics{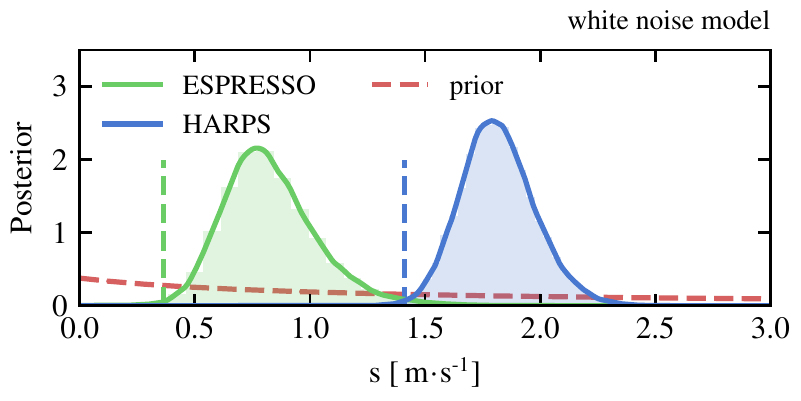}\\
        \includegraphics{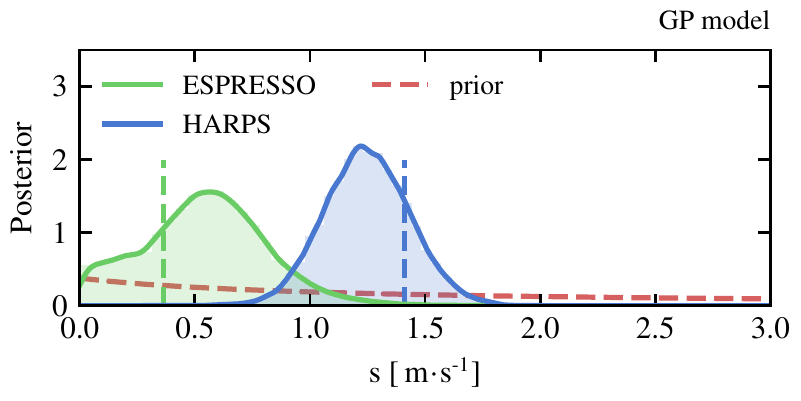}\\
        \includegraphics{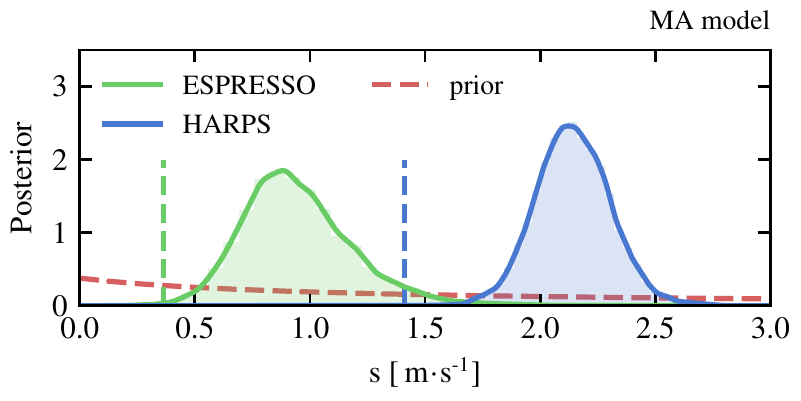}
        \caption{Posterior distributions for the additional white noise parameters
          from the white noise model (top), the GP model (middle), and the MA
          model (bottom). The distributions for HARPS and ESPRESSO are shown in
          the same colours as in \fig{fig:RVs}. Dashed vertical lines correspond to
          the mean RV uncertainty for each instrument.}
        \label{fig:jitters}
      \end{figure}

      From a purely statistical point of view, the evidence for the GP model ($\ln
      Z=-606.32$) is significantly higher than for both the
      white noise ($\ln Z=-622.35$) or the MA ($\ln
      Z=-623.33$) models. Even though the three models have a
      different number of free parameters and, thus, different parameter spaces,
      the value of the evidence can be trusted to provide the most meaningful
      comparison given the observed data and our priors.

      Moreover, the parameters in the GP model can be readily interpreted in
      physical terms. Our results for all four GP hyperparameters (see
      \fig{fig:posteriors_etas}) agree with the physical interpretation of a
      rotating star ($P_{\rm rot} = 25.62$ days) with
      active regions going in and out of view ($\eta_4 =
      0.68$) over a lifetime of about two rotation
      periods ($\sim\!42.55$ days), and an
      activity-induced RV amplitude that reproduces the observed data. The value
      for the rotation period matches closely the spectroscopic determination
      from $v \sin i$ and the estimate from the \TESS photometry (see Appendix
      \ref{app:tess}). We also note that our estimate for $\eta_2$ corresponds to
      an ‘effective' lifetime of both spots and faculae. Its relatively
      large uncertainty may be caused in part by changes in the real lifetime
      of active regions over the 15 years of observations.
      
      In terms of the RV residuals, our results also point to the GP model being
      the best description of the data. Figure \ref{fig:jitters} shows the
      posterior distributions for the additional white noise parameters for each
      of the two instruments and each of the three models. These parameters
      reflect the amount of RV variation observed in the data that is not
      accommodated by other parts of the models and that must, therefore, be attributed
      to white, uncorrelated noise. The posteriors are only compatible with the
      average RV uncertainties (shown as dashed lines in \fig{fig:jitters}) in the
      case of the GP model. The results for the additional white noise of ESPRESSO
      are consistent with the fact that the uncertainties only contain the photon
      noise contribution and are thus slightly underestimated (by about 18 $\cms$,
      on average).

      Regarding the white noise and MA models, we argue that the activity-induced
      RV signal present in the data is not well modelled with these two noise
      models. The white noise model can only separate RV variations into Keplerian
      or white noise components, which leads to a large number of detected signals
      as well as a high residual rms (see \fig{fig:phasek3}). The moving average
      term included in the MA model does not seem to be able to account for RV
      variations over the timescales related to stellar activity, and can only
      explain a very small part of the observed RV variations. For these reasons,
      we consider that the results regarding the number and the parameters of the
      Keplerian signals from these two models are not reliable.

    \subsection{Detectability limits}

      \begin{figure}
        \centering
        \includegraphics[width=\hsize]{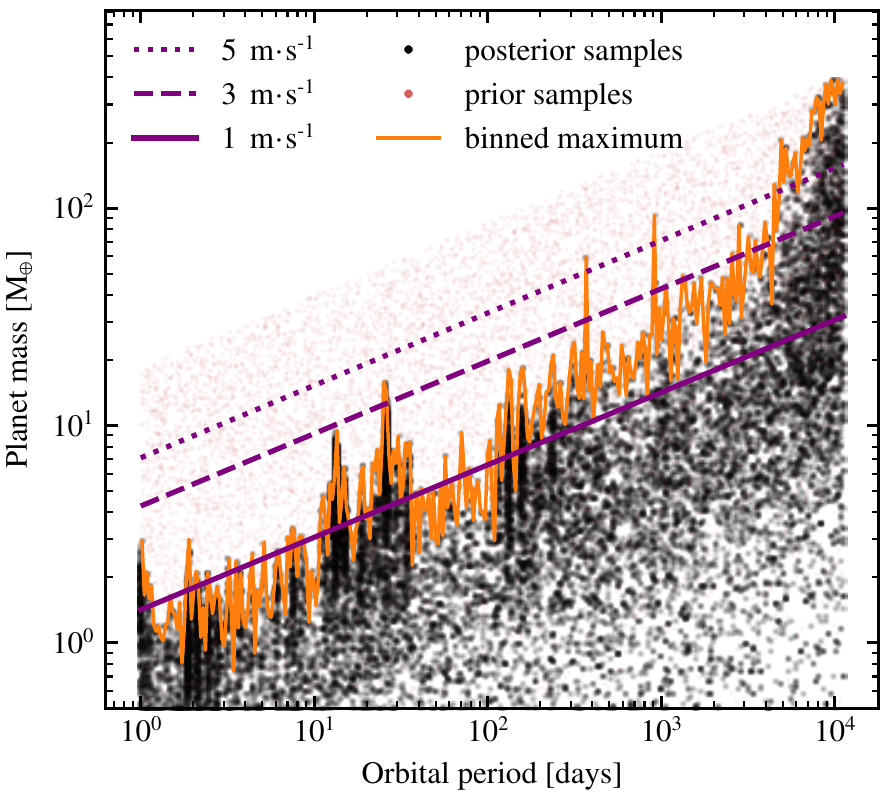}
        \caption{Detectability limits from the analysis with the GP model. Black
          points show posterior samples with $\np \ge 1$ and the orange curve
          represents the maximum planet mass from those samples, in 300 bins in
          log-period. Red points are samples from the prior distribution and the
          diagonal lines correspond to signals with amplitudes of 5, 3, and 1
          $\ms$.}
        \label{fig:detlim}
      \end{figure}

      Accepting the GP model as the best explanation for the observed RV
      variations of \target leads to the detection of zero planetary companions.
      However, the currently available data are not sensitive to every possible
      planet signal. We can use the posterior distributions obtained in our
      analysis to estimate which planets, in terms of their minimum masses and
      orbital periods, could still be compatible with the current data and
      detectable with further future observations.

      If we assume that the posterior distribution is sufficiently well
      explored, the sampler visits the regions of parameter space that
      contribute substantially to the posterior. Therefore, the resulting
      samples cover the areas of parameter space which are allowed by the priors
      and have relatively high likelihood (that is, they are compatible with the
      data). For samples which have $\np>\np^{\rm detected}$, this means that
      the sampler visits the highest possible Keplerian amplitudes allowed by
      the data, as well as amplitudes so low that the data cannot rule them out.
      The former provide a detectability limit%
      \footnote{We deliberately avoid the term \emph{detection} limit, because
      there is no planet detection.}.
      We note that these limits include the GP stellar activity model.

      From all the posterior samples of the GP model, we select those with $\np
      \ge 1$. With the orbital periods, semi-amplitudes, and eccentricities of
      these samples (for each of the up to four Keplerians), and the stellar
      mass given in Table \ref{tab:parameters}, we calculate the minimum mass a
      planet with those parameters would have. Figure \ref{fig:detlim} shows all
      the samples in the planet mass -- orbital period space. The maximum mass
      at each period corresponds to the detectability limit with the current
      data. Samples from the prior distribution are also shown. 
      
      The available data can constrain the posterior at shorter orbital periods
      but less so for periods longer than the time span of observations
      (5615
days). The increase in the detectability limits
      in the region between ten and 40 days is due to the effect of stellar
      activity, which can conceal more massive planets that would otherwise be
      detected if stellar activity was not present. The increase in the average
      level of the detectability limits after orbital periods around 100 days is
      due to the time span of the ESPRESSO observations
      (128
days) after which the posterior is
      constrained mostly by the HARPS data, which have a lower precision.

\section{Conclusions}
\label{sec:7}

    We obtained ESPRESSO observations of \target with the goal of confirming
    or disproving the presence of planetary companions around this star. The
    ESPRESSO RVs show a much improved precision when compared to previous
    observations from HARPS. Analysis of the full data set with different noise
    models leads to different conclusions regarding the number of significant
    Keplerian signals detected in the data. 
    
    We conclude that the GP model best explains the observed RV variations because
    \emph{(i)} it has a higher evidence when compared to the other models;
    \emph{(ii)} it allows for the stellar rotation period and active region
    evolution timescale to be determined; and \emph{(iii)} the level of the RV
    residuals and model jitters agree with the estimated uncertainties for each
    instrument.
    
    Analysis of the HARPS activity indicators shows clear evidence pointing to the
    chromospheric origin of the RV variations: some indicators show periodicities
    in common with the RVs, and the FWHM of the HARPS CCF in particular shows a
    clear time lag of $\sim\!3$ days relative to the RVs, which is known to be a
    sign of stellar activity \citep{CollierCameron2019}. The ESPRESSO activity
    indicators, on the other hand, show less clear variations and periodicities,
    which we attribute to a quieter stellar activity phase during the ESPRESSO
    observations.

    Overall, \target is a telling example of the difficulties introduced by
    stellar activity when trying to detect planets with RV observations. The
    presence of several (high amplitude) signals in the RV time series is not up
    for discussion but their interpretation with relation to orbiting planets is
    complex. Confirming the presence of a planet is often much harder than
    simply finding a significant periodic signal. Our conclusion from the
    currently available data is that they do not provide enough evidence for the
    detection of a planet and our stellar activity model is both physically
    interpretable and statistically preferred.

    This work presents, for the first time, radial velocities obtained with
    ESPRESSO. The data clearly show that the instrument is already delivering on
    the expected RV precision in its first semester of operations. The limited
    time span of our observations does not allow us to comment on the long-term
    RV stability but the gain in precision relative to current state-of-the-art
    spectrographs like HARPS is evident.

    In the future, it will be important to develop new activity indicators which
    can track stellar activity down to this unprecedented $\cms$ level
    \citep[see, e.g.][]{Lanza2018,Wise2018}. Moreover, extracting RVs using
    spectral lines less sensitive to activity \citep[e.g.][]{Dumusque2018}, or
    joint modelling of the RVs and some activity indicators
    \citep{Rajpaul2015,Jones2017} may also help in disentangling stellar
    activity from planets.

  \begin{acknowledgements}
    The authors of this paper are ordered alphabetically \mbox{after} the first
    author. 
    J.P.F. led the ESO observing proposal, organised and performed all the
    analysis of the RV data, and wrote most of the manuscript. 
    S.G.S., V.A., N.C.S., A.M., and A.R.S. derived the stellar atmospheric
    parameters, $v\sin i$, and elemental abundances.
    S.C.C.B., O.D., M.O., and \mbox{E.M.A-G.} analysed the \TESS data.
    J.G.S. derived activity indices.
    P.F., S.U-M., P.T.P.V., and J.C. contributed to the data reduction and
    suggested statistical analyses.
    All authors contributed to discussions regarding acquisition, analysis, and
    interpretation of the data and were given the opportunity to review the
    results and comment on the manuscript.
    %
    We acknowledge the support from Fundação para a Ciência e Tecnologia (FCT,
    Portugal). In particular, this work was supported by FCT/MCTES through
    national funds and by FEDER-Fundo Europeu de Desenvolvimento Regional
    through COMPETE2020-Programa Operacional Competitividade e
    Internacionalização by these grants: UID/FIS/04434/2019,
    PTDC/FIS-AST/32113/2017 \& POCI-01-0145-FEDER-032113 and
    PTDC/FIS-AST/28953/2017 \& POCI-01-0145-FEDER-028953.
    J.P.F. and O.D. are supported in the form of work contracts funded by
    national funds through FCT with the references: DL 57/2016/CP1364/CT0005 and
    DL 57/2016/CP1364/CT0004.
    V.A., S.C.C.B., and S.G.S. also acknowledge support from FCT through
    Investigador FCT contracts: IF/00650/2015/CP1273/CT0001,
    IF/01312/2014/CP1215/CT0004 and IF/00028/2014/CP1215/CT0002.
    %
    %
    M.O. acknowledges the support of the DFG priority program SPP 1992
    ``Exploring the Diversity of Extrasolar Planets (RE 1664/17-1)''.
    We also acknowledge the support of the FCT/DAAD bilateral grant 2019 (DAAD
    ID: 57453096).
    This work has made use of the SIMBAD database, operated at CDS, Strasbourg,
    France and of data from the European Space Agency (ESA) mission {\it Gaia}
    (\url{https://www.cosmos.esa.int/gaia}), processed by the {\it Gaia} Data
    Processing and Analysis Consortium (DPAC,
    \url{https://www.cosmos.esa.int/web/gaia/dpac/consortium}). Funding for the
    DPAC has been provided by national institutions, in particular the
    institutions participating in the {\it Gaia} Multilateral Agreement. 
    This research made use of Lightkurve, a Python package for Kepler and TESS
    data analysis \citep{lightkurve}.

  \end{acknowledgements}

\bibliographystyle{aa}
\bibliography{references}

\begin{thebibliography}{83}
\expandafter\ifx\csname natexlab\endcsname\relax\def\natexlab#1{#1}\fi

\bibitem[{Adibekyan {et~al.}(2015)Adibekyan, Figueira, Santos, Sousa, Faria,
  {Delgado-Mena}, Oshagh, Tsantaki, Hakobyan, Gonz{\'a}lez~Hern{\'a}ndez,
  {Su{\'a}rez-Andr{\'e}s}, \& Israelian}]{Adibekyan2015a}
Adibekyan, V., Figueira, P., Santos, N.~C., {et~al.} 2015, A\&A, 583, A94

\bibitem[{Adibekyan {et~al.}(2012)Adibekyan, Delgado~Mena, Sousa, Santos,
  Israelian, Gonz{\'a}lez~Hern{\'a}ndez, Mayor, \& Hakobyan}]{Adibekyan2012}
Adibekyan, V.~Z., Delgado~Mena, E., Sousa, S.~G., {et~al.} 2012, A\&A, 547, A36

\bibitem[{{Astropy Collaboration} {et~al.}(2018){Astropy Collaboration},
  {Price-Whelan}, Sip{\H o}cz, G{\"u}nther, Lim, Crawford, Conseil, Shupe,
  Craig, Dencheva, Ginsburg, VanderPlas, Bradley, {P{\'e}rez-Su{\'a}rez}, {de
  Val-Borro}, Paper~Contributors, Aldcroft, Cruz, Robitaille, Tollerud,
  Coordination~Committee, Ardelean, Babej, Bach, Bachetti, Bakanov, Bamford,
  Barentsen, Barmby, Baumbach, Berry, Biscani, Boquien, Bostroem, Bouma,
  Brammer, Bray, Breytenbach, Buddelmeijer, Burke, Calderone,
  Cano~Rod{\'r}\i{}guez, Cara, Cardoso, Cheedella, Copin, Corrales, Crichton,
  D, Deil, Depagne, Dietrich, Donath, Droettboom, Earl, Erben, Fabbro,
  Ferreira, Finethy, Fox, Garrison, Gibbons, Goldstein, Gommers, Greco,
  Greenfield, Groener, Grollier, Hagen, Hirst, Homeier, Horton, Hosseinzadeh,
  Hu, Hunkeler, Ivezi{\'c}, Jain, Jenness, Kanarek, Kendrew, Kern, Kerzendorf,
  Khvalko, King, Kirkby, Kulkarni, Kumar, Lee, Lenz, Littlefair, Ma, Macleod,
  Mastropietro, McCully, Montagnac, Morris, Mueller, Mumford, Muna, Murphy,
  Nelson, Nguyen, Ninan, N{\"o}the, Ogaz, Oh, Parejko, Parley, Pascual, Patil,
  Patil, Plunkett, Prochaska, Rastogi, Reddy~Janga, Sabater, Sakurikar,
  Seifert, Sherbert, {Sherwood-Taylor}, Shih, Sick, Silbiger, Singanamalla,
  Singer, Sladen, Sooley, Sornarajah, Streicher, Teuben, Thomas, Tremblay,
  Turner, Terr{\'o}n, {van Kerkwijk}, {de la Vega}, Watkins, Weaver, Whitmore,
  Woillez, Zabalza, \& Contributors}]{astropy:2018}
{Astropy Collaboration}, {Price-Whelan}, A.~M., Sip{\H o}cz, B.~M., {et~al.}
  2018, AJ, 156, 123

\bibitem[{{Bailer-Jones} {et~al.}(2018){Bailer-Jones}, Rybizki, Fouesneau,
  Mantelet, \& Andrae}]{BailerJones2018}
{Bailer-Jones}, C. a.~L., Rybizki, J., Fouesneau, M., Mantelet, G., \& Andrae,
  R. 2018, AJ, 156, 58

\bibitem[{Barros {et~al.}(2016)Barros, Demangeon, \& Deleuil}]{Barros2016}
Barros, S. C.~C., Demangeon, O., \& Deleuil, M. 2016, A\&A, 594, A100

\bibitem[{{Bertran de Lis} {et~al.}(2015){Bertran de Lis}, Mena, Adibekyan,
  Santos, \& Sousa}]{BertrandeLis2015}
{Bertran de Lis}, S., Mena, E.~D., Adibekyan, V.~Z., Santos, N.~C., \& Sousa,
  S.~G. 2015, A\&A, 576, A89

\bibitem[{Boisse {et~al.}(2011)Boisse, Bouchy, H{\'e}brard, Bonfils, Santos, \&
  Vauclair}]{Boisse2011}
Boisse, I., Bouchy, F., H{\'e}brard, G., {et~al.} 2011, A\&A, 528, A4

\bibitem[{Bouchy {et~al.}(2001)Bouchy, Pepe, \& Queloz}]{Bouchy2001}
Bouchy, F., Pepe, F., \& Queloz, D. 2001, A\&A, 374, 733

\bibitem[{Brewer {et~al.}(2011)Brewer, P{\'a}rtay, \& Cs{\'a}nyi}]{Brewer2011}
Brewer, B.~J., P{\'a}rtay, L.~B., \& Cs{\'a}nyi, G. 2011, Statistics and
  Computing, 21, 649

\bibitem[{Chaplin {et~al.}(2019)Chaplin, Cegla, Watson, Davies, \&
  Ball}]{Chaplin2019}
Chaplin, W.~J., Cegla, H.~M., Watson, C.~A., Davies, G.~R., \& Ball, W.~H.
  2019, AJ, 157, 163

\bibitem[{Cloutier {et~al.}(2017)Cloutier, {Astudillo-Defru}, Doyon, Bonfils,
  Almenara, Benneke, Bouchy, Delfosse, Ehrenreich, Forveille, Lovis, Mayor,
  Menou, Murgas, Pepe, Rowe, Santos, Udry, \& W{\"u}nsche}]{Cloutier2017}
Cloutier, R., {Astudillo-Defru}, N., Doyon, R., {et~al.} 2017, A\&A, 608, A35

\bibitem[{Collier~Cameron {et~al.}(2019)Collier~Cameron, Mortier, Phillips,
  Dumusque, Haywood, Langellier, Watson, Cegla, Costes, Charbonneau, Coffinet,
  Latham, {Lopez-Morales}, Malavolta, Maldonado, Micela, Milbourne, Molinari,
  Saar, Thompson, Buchschacher, Cecconi, Cosentino, Ghedina, Glenday, Gonzalez,
  Li, Lodi, Lovis, Pepe, Poretti, Rice, Sasselov, Sozzetti, Szentgyorgyi, Udry,
  \& Walsworth}]{CollierCameron2019}
Collier~Cameron, A., Mortier, A., Phillips, D., {et~al.} 2019, MNRAS, 487, 1082

\bibitem[{Delgado~Mena {et~al.}(2010)Delgado~Mena, Israelian,
  Gonz{\'a}lez~Hern{\'a}ndez, Bond, Santos, Udry, \& Mayor}]{DelgadoMena2010}
Delgado~Mena, E., Israelian, G., Gonz{\'a}lez~Hern{\'a}ndez, J.~I., {et~al.}
  2010, ApJ, 725, 2349

\bibitem[{D{\'i}az {et~al.}(2018)D{\'i}az, Jenkins, Tuomi, Butler, Soto, Teske,
  Feng, Shectman, Arriagada, Crane, Thompson, \& Vogt}]{Diaz2018a}
D{\'i}az, M.~R., Jenkins, J.~S., Tuomi, M., {et~al.} 2018, AJ, 155, 126

\bibitem[{D{\'i}az {et~al.}(2016)D{\'i}az, S{\'e}gransan, Udry, Lovis, Pepe,
  Dumusque, Marmier, Alonso, Benz, Bouchy, Coffinet, Cameron, Deleuil,
  Figueira, Gillon, Curto, Mayor, Mordasini, Motalebi, Moutou, Pollacco,
  Pompei, Queloz, Santos, \& Wyttenbach}]{Diaz2016a}
D{\'i}az, R.~F., S{\'e}gransan, D., Udry, S., {et~al.} 2016, A\&A, 585, A134

\bibitem[{Dotter {et~al.}(2008)Dotter, Chaboyer, Jevremovi{\'c}, Kostov, Baron,
  \& Ferguson}]{Dotter2008}
Dotter, A., Chaboyer, B., Jevremovi{\'c}, D., {et~al.} 2008, ApJS, 178, 89

\bibitem[{Dumusque(2018)}]{Dumusque2018}
Dumusque, X. 2018, A\&A, 620, A47

\bibitem[{Dumusque {et~al.}(2011)Dumusque, Udry, Lovis, Santos, \&
  Monteiro}]{Dumusque2011b}
Dumusque, X., Udry, S., Lovis, C., Santos, N.~C., \& Monteiro, M. J. P. F.~G.
  2011, A\&A, 525, A140

\bibitem[{Edelson \& Krolik(1988)}]{Edelson1988}
Edelson, R.~A. \& Krolik, J.~H. 1988, ApJ, 333, 646

\bibitem[{Efron \& Gous(2001)}]{Efron2001}
Efron, B. \& Gous, A. 2001, Lecture Notes-Monograph Series, 38, 208

\bibitem[{Faria {et~al.}(2016)Faria, Haywood, Brewer, Figueira, Oshagh,
  Santerne, \& Santos}]{Faria2016}
Faria, J.~P., Haywood, R.~D., Brewer, B.~J., {et~al.} 2016, A\&A, 588, A31

\bibitem[{Faria {et~al.}(2018)Faria, Santos, Figueira, \& Brewer}]{Faria2018}
Faria, J.~P., Santos, N.~C., Figueira, P., \& Brewer, B.~J. 2018, JOSS, 3, 487

\bibitem[{Feng {et~al.}(2017)Feng, Tuomi, \& Jones}]{Feng2017b}
Feng, F., Tuomi, M., \& Jones, H. R.~A. 2017, MNRAS, 470, 4794

\bibitem[{Feng {et~al.}(2016)Feng, Tuomi, Jones, Butler, \& Vogt}]{Feng2016}
Feng, F., Tuomi, M., Jones, H. R.~A., Butler, R.~P., \& Vogt, S. 2016, MNRAS,
  461, 2440

\bibitem[{Feroz {et~al.}(2011)Feroz, Balan, \& Hobson}]{Feroz2011}
Feroz, F., Balan, S.~T., \& Hobson, M.~P. 2011, MNRAS, 415, 3462

\bibitem[{Figueira {et~al.}(2010)Figueira, Marmier, Bonfils, {di Folco}, Udry,
  Santos, Lovis, M{\'e}gevand, Melo, Pepe, Queloz, S{\'e}gransan, Triaud, \&
  Viana~Almeida}]{Figueira2010}
Figueira, P., Marmier, M., Bonfils, X., {et~al.} 2010, A\&A, 513, L8

\bibitem[{Figueira {et~al.}(2013)Figueira, Santos, Pepe, Lovis, \&
  Nardetto}]{Figueira2013}
Figueira, P., Santos, N.~C., Pepe, F., Lovis, C., \& Nardetto, N. 2013, A\&A,
  557, A93

\bibitem[{Fischer {et~al.}(2016)Fischer, {Anglada-Escude}, Arriagada, Baluev,
  Bean, Bouchy, Buchhave, Carroll, Chakraborty, Crepp, Dawson, Diddams,
  Dumusque, Eastman, Endl, Figueira, Ford, {Daniel Foreman-Mackey}, Fournier,
  F{\H u}r{\'e}sz, Gaudi, Gregory, Grundahl, Hatzes, H{\'e}brard, Herrero,
  Hogg, Howard, Johnson, {Paul Jorden}, Jurgenson, Latham, Laughlin, Loredo,
  Lovis, {Suvrath Mahadevan}, McCracken, Pepe, Perez, Phillips, Plavchan, {Lisa
  Prato}, Quirrenbach, Reiners, Robertson, Santos, Sawyer, {Damien Segransan},
  Sozzetti, Steinmetz, Szentgyorgyi, Udry, Valenti, Wang, Wittenmyer, \&
  Wright}]{Fischer2016}
Fischer, D.~A., {Anglada-Escude}, G., Arriagada, P., {et~al.} 2016, PASP, 128,
  066001

\bibitem[{Ford(2006)}]{Ford2006}
Ford, E.~B. 2006, in New {{Horizons}} in {{Astronomy}}: {{Frank N}}. {{Bash
  Symposium}}, Vol. 352, 15

\bibitem[{{Gaia Collaboration} {et~al.}(2018){Gaia Collaboration}, Brown,
  Vallenari, Prusti, {de Bruijne}, Babusiaux, {Bailer-Jones}, Biermann, Evans,
  Eyer, Jansen, Jordi, Klioner, Lammers, Lindegren, Luri, Mignard, Panem,
  Pourbaix, Randich, Sartoretti, Siddiqui, Soubiran, {van Leeuwen}, Walton,
  Arenou, Bastian, Cropper, Drimmel, Katz, Lattanzi, Bakker, Cacciari,
  Casta{\~n}eda, Chaoul, Cheek, De~Angeli, Fabricius, Guerra, Holl, Masana,
  Messineo, Mowlavi, Nienartowicz, Panuzzo, Portell, Riello, Seabroke, Tanga,
  Th{\'e}venin, {Gracia-Abril}, Comoretto, {Garcia-Reinaldos}, Teyssier,
  Altmann, Andrae, Audard, {Bellas-Velidis}, Benson, Berthier, Blomme, Burgess,
  Busso, Carry, Cellino, Clementini, Clotet, Creevey, Davidson, De~Ridder,
  Delchambre, Dell'Oro, Ducourant, {Fern{\'a}ndez-Hern{\'a}ndez}, Fouesneau,
  Fr{\'e}mat, Galluccio, {Garc{\'i}a-Torres}, {Gonz{\'a}lez-N{\'u}{\~n}ez},
  {Gonz{\'a}lez-Vidal}, Gosset, Guy, Halbwachs, Hambly, Harrison,
  Hern{\'a}ndez, Hestroffer, Hodgkin, Hutton, Jasniewicz,
  {Jean-Antoine-Piccolo}, Jordan, Korn, {Krone-Martins}, Lanzafame, Lebzelter,
  L{\"o}ffler, Manteiga, Marrese, {Mart{\'i}n-Fleitas}, Moitinho, Mora,
  Muinonen, Osinde, Pancino, Pauwels, Petit, {Recio-Blanco}, Richards,
  Rimoldini, Robin, Sarro, Siopis, Smith, Sozzetti, S{\"u}veges, Torra, {van
  Reeven}, Abbas, Abreu~Aramburu, Accart, Aerts, Altavilla, {\'A}lvarez,
  Alvarez, Alves, Anderson, Andrei, Anglada~Varela, Antiche, Antoja, Arcay,
  Astraatmadja, Bach, Baker, {Balaguer-N{\'u}{\~n}ez}, Balm, Barache, Barata,
  Barbato, Barblan, Barklem, Barrado, Barros, Barstow,
  Bartholom{\'e}~Mu{\~n}oz, Bassilana, Becciani, Bellazzini, Berihuete,
  Bertone, Bianchi, Bienaym{\'e}, {Blanco-Cuaresma}, Boch, Boeche, Bombrun,
  Borrachero, Bossini, Bouquillon, Bourda, Bragaglia, Bramante, Breddels,
  Bressan, Brouillet, Br{\"u}semeister, Brugaletta, Bucciarelli, Burlacu,
  Busonero, Butkevich, Buzzi, Caffau, Cancelliere, Cannizzaro, {Cantat-Gaudin},
  Carballo, Carlucci, Carrasco, Casamiquela, Castellani, {Castro-Ginard},
  Charlot, Chemin, Chiavassa, Cocozza, Costigan, Cowell, Crifo, Crosta,
  Crowley, Cuypers\textdagger, Dafonte, Damerdji, Dapergolas, David, David, {de
  Laverny}, De~Luise, De~March, {de Martino}, {de Souza}, {de Torres},
  Debosscher, {del Pozo}, Delbo, Delgado, Delgado, Di~Matteo, Diakite, Diener,
  Distefano, Dolding, Drazinos, Dur{\'a}n, Edvardsson, Enke, Eriksson, Esquej,
  Eynard~Bontemps, Fabre, Fabrizio, Faigler, Falc{\~a}o, Farr{\`a}s~Casas,
  Federici, Fedorets, Fernique, Figueras, Filippi, Findeisen, Fonti, Fraile,
  Fraser, Fr{\'e}zouls, Gai, Galleti, Garabato, {Garc{\'i}a-Sedano}, Garofalo,
  Garralda, Gavel, Gavras, Gerssen, Geyer, Giacobbe, Gilmore, Girona,
  Giuffrida, Glass, Gomes, Granvik, Gueguen, Guerrier, Guiraud,
  {Guti{\'e}rrez-S{\'a}nchez}, Haigron, Hatzidimitriou, Hauser, Haywood,
  Heiter, Helmi, Heu, Hilger, Hobbs, Hofmann, Holland, Huckle, Hypki, Icardi,
  Jan{\ss}en, {Jevardat de Fombelle}, Jonker, Juh{\'a}sz, Julbe, Karampelas,
  Kewley, Klar, Kochoska, Kohley, Kolenberg, Kontizas, Kontizas, Koposov,
  Kordopatis, {Kostrzewa-Rutkowska}, Koubsky, Lambert, Lanza, Lasne, Lavigne,
  Le~Fustec, {Le Poncin-Lafitte}, Lebreton, Leccia, Leclerc, {Lecoeur-Taibi},
  Lenhardt, Leroux, Liao, Licata, Lindstr{\o}m, Lister, Livanou, Lobel,
  L{\'o}pez, Managau, Mann, Mantelet, Marchal, Marchant, Marconi, Marinoni,
  Marschalk{\'o}, Marshall, Martino, Marton, Mary, Massari, Matijevi{\v c},
  Mazeh, McMillan, Messina, Michalik, Millar, Molina, Molinaro, Moln{\'a}r,
  Montegriffo, Mor, Morbidelli, Morel, Morris, Mulone, Muraveva, Musella,
  Nelemans, Nicastro, Noval, O'Mullane, Ord{\'e}novic,
  {Ord{\'o}{\~n}ez-Blanco}, Osborne, Pagani, Pagano, Pailler, Palacin,
  Palaversa, Panahi, Pawlak, Piersimoni, Pineau, Plachy, Plum, Poggio,
  Poujoulet, Pr{\v s}a, Pulone, Racero, Ragaini, Rambaux, {Ramos-Lerate},
  Regibo, Reyl{\'e}, Riclet, Ripepi, Riva, Rivard, Rixon, Roegiers, Roelens,
  {Romero-G{\'o}mez}, Rowell, Royer, {Ruiz-Dern}, Sadowski,
  Sagrist{\`a}~Sell{\'e}s, Sahlmann, Salgado, Salguero, Sanna, {Santana-Ros},
  Sarasso, Savietto, Schultheis, Sciacca, Segol, Segovia, S{\'e}gransan, Shih,
  Siltala, Silva, Smart, Smith, Solano, Solitro, Sordo, Soria~Nieto, Souchay,
  Spagna, Spoto, Stampa, Steele, Steidelm{\"u}ller, Stephenson, Stoev, Suess,
  Surdej, Szabados, {Szegedi-Elek}, Tapiador, Taris, Tauran, Taylor, Teixeira,
  Terrett, Teyssandier, Thuillot, Titarenko, Torra~Clotet, Turon, Ulla,
  Utrilla, Uzzi, Vaillant, Valentini, Valette, {van Elteren}, Van~Hemelryck,
  {van Leeuwen}, Vaschetto, Vecchiato, Veljanoski, Viala, Vicente, Vogt, {von
  Essen}, Voss, Votruba, Voutsinas, Walmsley, Weiler, Wertz, Wevers,
  Wyrzykowski, Yoldas, {\v Z}erjal, Ziaeepour, Zorec, Zschocke, Zucker,
  Zurbach, \& Zwitter}]{GaiaCollaboration2018a}
{Gaia Collaboration}, Brown, A. G.~A., Vallenari, A., {et~al.} 2018, A\&A, 616,
  A1

\bibitem[{{Gaia Collaboration} {et~al.}(2016){Gaia Collaboration}, Prusti, {de
  Bruijne}, Brown, Vallenari, Babusiaux, {Bailer-Jones}, Bastian, Biermann,
  Evans, Eyer, Jansen, Jordi, Klioner, Lammers, Lindegren, Luri, Mignard,
  Milligan, Panem, Poinsignon, Pourbaix, Randich, Sarri, Sartoretti, Siddiqui,
  Soubiran, Valette, {van Leeuwen}, Walton, Aerts, Arenou, Cropper, Drimmel,
  H{\o}g, Katz, Lattanzi, O'Mullane, Grebel, Holland, Huc, Passot, Bramante,
  Cacciari, Casta{\~n}eda, Chaoul, Cheek, De~Angeli, Fabricius, Guerra,
  Hern{\'a}ndez, {Jean-Antoine-Piccolo}, Masana, Messineo, Mowlavi,
  Nienartowicz, {Ord{\'o}{\~n}ez-Blanco}, Panuzzo, Portell, Richards, Riello,
  Seabroke, Tanga, Th{\'e}venin, Torra, Els, {Gracia-Abril}, Comoretto,
  {Garcia-Reinaldos}, Lock, Mercier, Altmann, Andrae, Astraatmadja,
  {Bellas-Velidis}, Benson, Berthier, Blomme, Busso, Carry, Cellino,
  Clementini, Cowell, Creevey, Cuypers, Davidson, De~Ridder, {de Torres},
  Delchambre, Dell'Oro, Ducourant, Fr{\'e}mat, {Garc{\'i}a-Torres}, Gosset,
  Halbwachs, Hambly, Harrison, Hauser, Hestroffer, Hodgkin, Huckle, Hutton,
  Jasniewicz, Jordan, Kontizas, Korn, Lanzafame, Manteiga, Moitinho, Muinonen,
  Osinde, Pancino, Pauwels, Petit, {Recio-Blanco}, Robin, Sarro, Siopis, Smith,
  Smith, Sozzetti, Thuillot, {van Reeven}, Viala, Abbas, Abreu~Aramburu,
  Accart, Aguado, Allan, Allasia, Altavilla, {\'A}lvarez, Alves, Anderson,
  Andrei, Anglada~Varela, Antiche, Antoja, Ant{\'o}n, Arcay, Atzei, Ayache,
  Bach, Baker, {Balaguer-N{\'u}{\~n}ez}, Barache, Barata, Barbier, Barblan,
  Baroni, {Barrado y Navascu{\'e}s}, Barros, Barstow, Becciani, Bellazzini,
  Bellei, Bello~Garc{\'i}a, Belokurov, Bendjoya, Berihuete, Bianchi,
  Bienaym{\'e}, Billebaud, Blagorodnova, {Blanco-Cuaresma}, Boch, Bombrun,
  Borrachero, Bouquillon, Bourda, Bouy, Bragaglia, Breddels, Brouillet,
  Br{\"u}semeister, Bucciarelli, Budnik, Burgess, Burgon, Burlacu, Busonero,
  Buzzi, Caffau, Cambras, Campbell, Cancelliere, {Cantat-Gaudin}, Carlucci,
  Carrasco, Castellani, Charlot, Charnas, Charvet, Chassat, Chiavassa, Clotet,
  Cocozza, Collins, Collins, Costigan, Crifo, Cross, Crosta, Crowley, Dafonte,
  Damerdji, Dapergolas, David, David, De~Cat, {de Felice}, {de Laverny},
  De~Luise, De~March, {de Martino}, {de Souza}, Debosscher, {del Pozo}, Delbo,
  Delgado, Delgado, {di Marco}, Di~Matteo, Diakite, Distefano, Dolding,
  Dos~Anjos, Drazinos, Dur{\'a}n, Dzigan, Ecale, Edvardsson, Enke, Erdmann,
  Escolar, Espina, Evans, Eynard~Bontemps, Fabre, Fabrizio, Faigler,
  Falc{\~a}o, Farr{\`a}s~Casas, Faye, Federici, Fedorets,
  {Fern{\'a}ndez-Hern{\'a}ndez}, Fernique, Fienga, Figueras, Filippi,
  Findeisen, Fonti, Fouesneau, Fraile, Fraser, Fuchs, Furnell, Gai, Galleti,
  Galluccio, Garabato, {Garc{\'i}a-Sedano}, Gar{\'e}, Garofalo, Garralda,
  Gavras, Gerssen, Geyer, Gilmore, Girona, Giuffrida, Gomes,
  {Gonz{\'a}lez-Marcos}, {Gonz{\'a}lez-N{\'u}{\~n}ez}, {Gonz{\'a}lez-Vidal},
  Granvik, Guerrier, Guillout, Guiraud, G{\'u}rpide,
  {Guti{\'e}rrez-S{\'a}nchez}, Guy, Haigron, Hatzidimitriou, Haywood, Heiter,
  Helmi, Hobbs, Hofmann, Holl, Holland, Hunt, Hypki, Icardi, Irwin, {Jevardat
  de Fombelle}, Jofr{\'e}, Jonker, Jorissen, Julbe, Karampelas, Kochoska,
  Kohley, Kolenberg, Kontizas, Koposov, Kordopatis, Koubsky, Kowalczyk,
  {Krone-Martins}, Kudryashova, Kull, Bachchan, {Lacoste-Seris}, Lanza,
  Lavigne, {Le Poncin-Lafitte}, Lebreton, Lebzelter, Leccia, Leclerc,
  {Lecoeur-Taibi}, Lemaitre, Lenhardt, Leroux, Liao, Licata, Lindstr{\o}m,
  Lister, Livanou, Lobel, L{\"o}ffler, L{\'o}pez, {Lopez-Lozano}, Lorenz,
  Loureiro, MacDonald, Magalh{\~a}es~Fernandes, Managau, Mann, Mantelet,
  Marchal, Marchant, Marconi, Marie, Marinoni, Marrese, Marschalk{\'o},
  Marshall, {Mart{\'i}n-Fleitas}, Martino, Mary, Matijevi{\v c}, Mazeh,
  McMillan, Messina, Mestre, Michalik, Millar, Miranda, Molina, Molinaro,
  Molinaro, Moln{\'a}r, Moniez, Montegriffo, Monteiro, Mor, Mora, Morbidelli,
  Morel, Morgenthaler, Morley, Morris, Mulone, Muraveva, Musella, Narbonne,
  Nelemans, Nicastro, Noval, Ord{\'e}novic, {Ordieres-Mer{\'e}}, Osborne,
  Pagani, Pagano, Pailler, Palacin, Palaversa, Parsons, Paulsen, Pecoraro,
  Pedrosa, Pentik{\"a}inen, Pereira, Pichon, Piersimoni, Pineau, Plachy, Plum,
  Poujoulet, Pr{\v s}a, Pulone, Ragaini, Rago, Rambaux, {Ramos-Lerate},
  Ranalli, Rauw, Read, Regibo, Renk, Reyl{\'e}, Ribeiro, Rimoldini, Ripepi,
  Riva, Rixon, Roelens, {Romero-G{\'o}mez}, Rowell, Royer, Rudolph,
  {Ruiz-Dern}, Sadowski, Sagrist{\`a}~Sell{\'e}s, Sahlmann, Salgado, Salguero,
  Sarasso, Savietto, Schnorhk, Schultheis, Sciacca, Segol, Segovia, Segransan,
  Serpell, Shih, Smareglia, Smart, Smith, Solano, Solitro, Sordo, Soria~Nieto,
  Souchay, Spagna, Spoto, Stampa, Steele, Steidelm{\"u}ller, Stephenson, Stoev,
  Suess, S{\"u}veges, Surdej, Szabados, {Szegedi-Elek}, Tapiador, Taris,
  Tauran, Taylor, Teixeira, Terrett, Tingley, Trager, Turon, Ulla, Utrilla,
  Valentini, {van Elteren}, Van~Hemelryck, {van Leeuwen}, Varadi, Vecchiato,
  Veljanoski, Via, Vicente, Vogt, Voss, Votruba, Voutsinas, Walmsley, Weiler,
  Weingrill, Werner, Wevers, Whitehead, Wyrzykowski, Yoldas, {\v Z}erjal,
  Zucker, Zurbach, Zwitter, Alecu, Allen, Allende~Prieto, Amorim,
  {Anglada-Escud{\'e}}, Arsenijevic, Azaz, Balm, Beck, Bernstein, Bigot,
  Bijaoui, Blasco, Bonfigli, Bono, Boudreault, Bressan, Brown, Brunet,
  Bunclark, Buonanno, Butkevich, Carret, Carrion, Chemin, Ch{\'e}reau,
  Corcione, Darmigny, {de Boer}, {de Teodoro}, {de Zeeuw}, Delle~Luche,
  Domingues, Dubath, Fodor, Fr{\'e}zouls, Fries, Fustes, Fyfe, Gallardo,
  Gallegos, Gardiol, Gebran, Gomboc, G{\'o}mez, Grux, Gueguen, Heyrovsky, Hoar,
  Iannicola, Isasi~Parache, Janotto, Joliet, Jonckheere, Keil, Kim, Klagyivik,
  Klar, Knude, Kochukhov, Kolka, Kos, Kutka, Lainey, LeBouquin, Liu, Loreggia,
  Makarov, Marseille, Martayan, {Martinez-Rubi}, Massart, Meynadier, Mignot,
  Munari, Nguyen, Nordlander, Ocvirk, O'Flaherty, Olias~Sanz, Ortiz, Osorio,
  Oszkiewicz, Ouzounis, Palmer, Park, Pasquato, Peltzer, Peralta, P{\'e}turaud,
  Pieniluoma, Pigozzi, Poels, Prat, Prod'homme, Raison, Rebordao, Risquez,
  {Rocca-Volmerange}, Rosen, {Ruiz-Fuertes}, Russo, Sembay, Serraller~Vizcaino,
  Short, Siebert, Silva, Sinachopoulos, Slezak, Soffel, Sosnowska, Strai{\v
  z}ys, {ter Linden}, Terrell, Theil, Tiede, Troisi, Tsalmantza, Tur, Vaccari,
  Vachier, Valles, Van~Hamme, Veltz, Virtanen, Wallut, Wichmann, Wilkinson,
  Ziaeepour, \& Zschocke}]{GaiaCollaboration2016a}
{Gaia Collaboration}, Prusti, T., {de Bruijne}, J. H.~J., {et~al.} 2016, A\&A,
  595, A1

\bibitem[{Giles {et~al.}(2017)Giles, Collier~Cameron, \& Haywood}]{Giles2017}
Giles, H. A.~C., Collier~Cameron, A., \& Haywood, R.~D. 2017, MNRAS, 472, 1618

\bibitem[{{Gomes da Silva} {et~al.}(2018){Gomes da Silva}, Figueira, Santos, \&
  Faria}]{GomesdaSilva2018}
{Gomes da Silva}, J., Figueira, P., Santos, N.~C., \& Faria, J.~P. 2018, JOSS,
  3, 667

\bibitem[{{Gomes da Silva} {et~al.}(2011){Gomes da Silva}, Santos, Bonfils,
  Delfosse, Forveille, \& Udry}]{GomesdaSilva2011}
{Gomes da Silva}, J., Santos, N.~C., Bonfils, X., {et~al.} 2011, A\&A, 534, A30

\bibitem[{Gregory(2005)}]{Gregory2005}
Gregory, P.~C. 2005, ApJ, 631, 1198

\bibitem[{Gregory(2011)}]{Gregory2011}
Gregory, P.~C. 2011, MNRAS, 410, 94

\bibitem[{Haywood(2016)}]{Haywood2016a}
Haywood, R.~D. 2016, PhD thesis, University of St Andrews, {St Andrews}

\bibitem[{Haywood {et~al.}(2014)Haywood, Collier~Cameron, Queloz, Barros,
  Deleuil, Fares, Gillon, Lanza, Lovis, Moutou, Pepe, Pollacco, Santerne,
  S{\'e}gransan, \& Unruh}]{Haywood2014}
Haywood, R.~D., Collier~Cameron, A., Queloz, D., {et~al.} 2014, MNRAS, 443,
  2517

\bibitem[{Hippke \& Heller(2019)}]{Hippke2019}
Hippke, M. \& Heller, R. 2019, A\&A, 623, A39

\bibitem[{Jeffreys(1961)}]{Jeffreys1961}
Jeffreys, H. 1961, Theory of Probability, 3rd edn. ({Oxford}: {Clarendon
  Press})

\bibitem[{Jenkins {et~al.}(2017)Jenkins, Jones, Tuomi, D{\'i}az, Cordero,
  Aguayo, Pantoja, Arriagada, Mahu, Brahm, Rojo, Soto, Ivanyuk, Becerra~Yoma,
  {Day-Jones}, Ruiz, Pavlenko, Barnes, Murgas, Pinfield, Jones,
  {L{\'o}pez-Morales}, Shectman, Butler, \& Minniti}]{Jenkins2017}
Jenkins, J.~S., Jones, H. R.~A., Tuomi, M., {et~al.} 2017, MNRAS, 466, 443

\bibitem[{Jenkins \& Tuomi(2014)}]{Jenkins2014}
Jenkins, J.~S. \& Tuomi, M. 2014, ApJ, 794, 110

\bibitem[{Jenkins {et~al.}(2013)Jenkins, Tuomi, Brasser, Ivanyuk, \&
  Murgas}]{Jenkins2013}
Jenkins, J.~S., Tuomi, M., Brasser, R., Ivanyuk, O., \& Murgas, F. 2013, ApJ,
  771, 41

\bibitem[{Jones {et~al.}(2017)Jones, Stenning, Ford, Wolpert, Loredo, \&
  Dumusque}]{Jones2017}
Jones, D.~E., Stenning, D.~C., Ford, E.~B., {et~al.} 2017, ArXiv e-prints
  [\eprint[arXiv]{1711.01318}]

\bibitem[{Jones {et~al.}(2001)Jones, Oliphant, Peterson, {et~al.}}]{Jones2001}
Jones, E., Oliphant, T., Peterson, P., {et~al.} 2001, {{SciPy}}: {{Open}}
  Source Scientific Tools for {{Python}}, [online; accessed 19-04-2018]

\bibitem[{Kipping(2013)}]{Kipping2013}
Kipping, D.~M. 2013, Mon Not R Astron Soc Lett, 434, L51

\bibitem[{Kov{\'a}cs {et~al.}(2002)Kov{\'a}cs, Zucker, \& Mazeh}]{Kovacs2002}
Kov{\'a}cs, G., Zucker, S., \& Mazeh, T. 2002, A\&A, 391, 369

\bibitem[{Kumaraswamy(1980)}]{Kumaraswamy1980}
Kumaraswamy, P. 1980, Journal of Hydrology, 46, 79

\bibitem[{Kurucz(1993)}]{kurucz1993}
Kurucz, R.~L. 1993, {{SYNTHE}} Spectrum Synthesis Programs and Line Data

\bibitem[{Lanza {et~al.}(2018)Lanza, Malavolta, Benatti, Desidera, Bignamini,
  Bonomo, Esposito, Figueira, Gratton, Scandariato, Damasso, Sozzetti, Biazzo,
  Claudi, Cosentino, Covino, Maggio, Masiero, Micela, Molinari, Pagano, Piotto,
  Poretti, Smareglia, Affer, Boccato, Borsa, Boschin, Giacobbe, Knapic, Leto,
  Maldonado, Mancini, Fiorenzano, Messina, Nascimbeni, Pedani, \&
  Rainer}]{Lanza2018}
Lanza, A.~F., Malavolta, L., Benatti, S., {et~al.} 2018, A\&A, 616, A155

\bibitem[{{Lightkurve Collaboration} {et~al.}(2018){Lightkurve Collaboration},
  Cardoso, Hedges, {Gully-Santiago}, Saunders, Cody, Barclay, Hall, Sagear,
  Turtelboom, Zhang, Tzanidakis, Mighell, Coughlin, Bell, {Berta-Thompson},
  Williams, Dotson, \& Barentsen}]{lightkurve}
{Lightkurve Collaboration}, Cardoso, J. V. d.~M., Hedges, C., {et~al.} 2018,
  Astrophysics Source Code Library [\eprint{1812.013}]

\bibitem[{Lo~Curto {et~al.}(2015)Lo~Curto, Pepe, Avila, Boffin, Bovay,
  Chazelas, Coffinet, Fleury, Hughes, Lovis, Maire, Manescau, Pasquini, Rihs,
  Sinclaire, \& Udry}]{LoCurto2015}
Lo~Curto, G., Pepe, F., Avila, G., {et~al.} 2015, The Messenger, 162, 9

\bibitem[{{L{\'o}pez-Morales} {et~al.}(2016){L{\'o}pez-Morales}, Haywood,
  Coughlin, Zeng, Buchhave, Giles, Affer, Bonomo, Charbonneau, Cameron,
  {Rosario Consentino}, Dressing, Dumusque, Figueira, Fiorenzano, {Avet
  Harutyunyan}, Johnson, Latham, Lopez, Lovis, Malavolta, {Michel Mayor},
  Micela, Molinari, Mortier, Motalebi, Nascimbeni, {Francesco Pepe}, Phillips,
  Piotto, Pollacco, Queloz, Rice, Sasselov, {Damien Segransan}, Sozzetti, Udry,
  Vanderburg, \& Watson}]{Lopez-Morales2016}
{L{\'o}pez-Morales}, M., Haywood, R.~D., Coughlin, J.~L., {et~al.} 2016, AJ,
  152, 204

\bibitem[{Mandel \& Agol(2002)}]{Mandel2002}
Mandel, K. \& Agol, E. 2002, ApJ, 580, L171

\bibitem[{Mayor {et~al.}(2003)Mayor, Pepe, Queloz, Bouchy, Rupprecht, Lo~Curto,
  Avila, Benz, Bertaux, Bonfils, Dall, Dekker, Delabre, Eckert, Fleury,
  Gilliotte, Gojak, Guzman, Kohler, Lizon, Longinotti, Lovis, Megevand,
  Pasquini, Reyes, Sivan, Sosnowska, Soto, Udry, {van Kesteren}, Weber, \&
  Weilenmann}]{Mayor2003}
Mayor, M., Pepe, F., Queloz, D., {et~al.} 2003, The Messenger, 114, 20

\bibitem[{McQuillan {et~al.}(2014)McQuillan, Mazeh, \& Aigrain}]{McQuillan2014}
McQuillan, A., Mazeh, T., \& Aigrain, S. 2014, ApJS, 211, 24

\bibitem[{Melo {et~al.}(2007)Melo, Santos, Gieren, Pietrzynski, Ruiz, Sousa,
  Bouchy, Lovis, Mayor, Pepe, Queloz, {da Silva}, \& Udry}]{Melo2007}
Melo, C., Santos, N.~C., Gieren, W., {et~al.} 2007, A\&A, 467, 721

\bibitem[{Modigliani {et~al.}(2019)Modigliani, Sownsowska, \&
  Lovis}]{Modigliani2019}
Modigliani, A., Sownsowska, D., \& Lovis, C. 2019, {{ESPRESSO Pipeline User
  Manual}}, ESO

\bibitem[{Mortier {et~al.}(2015)Mortier, Faria, Correia, Santerne, \&
  Santos}]{Mortier2015}
Mortier, A., Faria, J.~P., Correia, C.~M., Santerne, A., \& Santos, N.~C. 2015,
  A\&A, 573, A101

\bibitem[{Morton(2015)}]{Morton2015}
Morton, T.~D. 2015, ASCL, ascl:1503.010

\bibitem[{Noyes {et~al.}(1984)Noyes, Hartmann, Baliunas, Duncan, \&
  Vaughan}]{Noyes1984}
Noyes, R.~W., Hartmann, L.~W., Baliunas, S.~L., Duncan, D.~K., \& Vaughan,
  A.~H. 1984, ApJ, 279, 763

\bibitem[{Pepe {et~al.}(2014)Pepe, Molaro, Cristiani, Rebolo, Santos, Dekker,
  M{\'e}gevand, Zerbi, Cabral, Di~Marcantonio, {et~al.}}]{Pepe2014}
Pepe, F., Molaro, P., Cristiani, S., {et~al.} 2014, Astron. Nachr., 335, 8

\bibitem[{Perryman(2014)}]{Perryman2014}
Perryman, M. A.~C. 2014, The {{Exoplanet Handbook}}, 1st edn. ({Cambridge
  University Press})

\bibitem[{Queloz {et~al.}(2001)Queloz, Henry, Sivan, Baliunas, Beuzit, Donahue,
  Mayor, Naef, Perrier, \& Udry}]{Queloz2001}
Queloz, D., Henry, G.~W., Sivan, J.~P., {et~al.} 2001, A\&A, 379, 279

\bibitem[{Rajpaul {et~al.}(2015)Rajpaul, Aigrain, Osborne, Reece, \&
  Roberts}]{Rajpaul2015}
Rajpaul, V., Aigrain, S., Osborne, M.~A., Reece, S., \& Roberts, S. 2015,
  MNRAS, 452, 2269

\bibitem[{Rasmussen \& Williams(2006)}]{Rasmussen2006}
Rasmussen, C.~E. \& Williams, C. K.~I. 2006, Gaussian Processes for Machine
  Learning ({Cambridge}: {MIT Press})

\bibitem[{Ricker {et~al.}(2015)Ricker, Winn, Vanderspek, Latham, Bakos, Bean,
  {Berta-Thompson}, Brown, Buchhave, Butler, Butler, Chaplin, Charbonneau,
  {Christensen-Dalsgaard}, Clampin, Deming, Doty, De~Lee, Dressing, Dunham,
  Endl, Fressin, Ge, Henning, Holman, Howard, Ida, Jenkins, Jernigan, Johnson,
  Kaltenegger, Kawai, Kjeldsen, Laughlin, Levine, Lin, Lissauer, MacQueen,
  Marcy, McCullough, Morton, Narita, Paegert, Palle, Pepe, Pepper, Quirrenbach,
  Rinehart, Sasselov, Sato, Seager, Sozzetti, Stassun, Sullivan, Szentgyorgyi,
  Torres, Udry, \& Villasenor}]{Ricker2015}
Ricker, G.~R., Winn, J.~N., Vanderspek, R., {et~al.} 2015, JATIS, 1, 014003

\bibitem[{Saar \& Donahue(1997)}]{Saar1997}
Saar, S.~H. \& Donahue, R.~A. 1997, ApJ, 485, 319

\bibitem[{Santos {et~al.}(2000)Santos, Mayor, Naef, Pepe, Queloz, Udry, \&
  Blecha}]{Santos2000}
Santos, N.~C., Mayor, M., Naef, D., {et~al.} 2000, A\&A, 361, 265

\bibitem[{Santos {et~al.}(2014)Santos, Mortier, Faria, Dumusque, Adibekyan,
  {Delgado-Mena}, Figueira, Benamati, Boisse, Cunha, {Gomes da Silva},
  Lo~Curto, Lovis, Martins, Mayor, Melo, Oshagh, Pepe, Queloz, Santerne,
  S{\'e}gransan, Sozzetti, Sousa, \& Udry}]{Santos2014}
Santos, N.~C., Mortier, A., Faria, J.~P., {et~al.} 2014, A\&A, 566, A35

\bibitem[{Santos {et~al.}(2013)Santos, Sousa, Mortier, Neves, Adibekyan,
  Tsantaki, Mena, Bonfils, Israelian, Mayor, \& Udry}]{Santos2013}
Santos, N.~C., Sousa, S.~G., Mortier, A., {et~al.} 2013, A\&A, 556, A150

\bibitem[{Shapiro {et~al.}(2019)Shapiro, {Amazo-G{\'o}mez}, Krivova, \&
  Solanki}]{Shapiro2019}
Shapiro, A., {Amazo-G{\'o}mez}, E., Krivova, N., \& Solanki, S. 2019, A\&A,
  submitted

\bibitem[{Shapiro {et~al.}(2017)Shapiro, Solanki, Krivova, Cameron, Yeo, \&
  Schmutz}]{Shapiro2017}
Shapiro, A.~I., Solanki, S.~K., Krivova, N.~A., {et~al.} 2017, Nature
  Astronomy, 1, 612

\bibitem[{Sneden {et~al.}(2012)Sneden, Bean, Ivans, Lucatello, \&
  Sobeck}]{Sneden2012}
Sneden, C., Bean, J., Ivans, I., Lucatello, S., \& Sobeck, J. 2012,
  Astrophysics Source Code Library, ascl:1202.009

\bibitem[{Solanki {et~al.}(2006)Solanki, Inhester, \&
  Sch{\"u}ssler}]{Solanki2006}
Solanki, S.~K., Inhester, B., \& Sch{\"u}ssler, M. 2006, Rep. Prog. Phys., 69,
  563

\bibitem[{Sousa(2014)}]{Sousa2014}
Sousa, S.~G. 2014, in Determination of {{Atmospheric Parameters}} of {{B}}-,
  {{A}}-, {{F}}- and {{G}}-{{Type Stars}}, ed. E.~Niemczura, B.~Smalley, \&
  W.~Pych ({Cham}: {Springer International Publishing}), 297--310

\bibitem[{Sousa {et~al.}(2007)Sousa, Santos, Israelian, Mayor, \&
  Monteiro}]{Sousa2007}
Sousa, S.~G., Santos, N.~C., Israelian, G., Mayor, M., \& Monteiro, M. J. P.
  F.~G. 2007, A\&A, 469, 783

\bibitem[{Sousa {et~al.}(2008)Sousa, Santos, Mayor, Udry, Casagrande,
  Israelian, Pepe, Queloz, \& Monteiro}]{Sousa2008}
Sousa, S.~G., Santos, N.~C., Mayor, M., {et~al.} 2008, A\&A, 487, 373

\bibitem[{Torrence \& Compo(1998)}]{Torrence1998}
Torrence, C. \& Compo, G.~P. 1998, Bulletin of the American Meteorological
  Society, 79, 61

\bibitem[{Torres {et~al.}(2010)Torres, Andersen, \& Gim{\'e}nez}]{Torres2010}
Torres, G., Andersen, J., \& Gim{\'e}nez, A. 2010, Astron Astrophys Rev, 18, 67

\bibitem[{Tuomi {et~al.}(2014)Tuomi, {Anglada-Escude}, Jenkins, \&
  Jones}]{Tuomi2014}
Tuomi, M., {Anglada-Escude}, G., Jenkins, J.~S., \& Jones, H. R.~A. 2014, ArXiv
  e-prints [\eprint[arXiv]{1405.2016}]

\bibitem[{Vaughan {et~al.}(1978)Vaughan, Preston, \& Wilson}]{Vaughan1978}
Vaughan, A.~H., Preston, G.~W., \& Wilson, O.~C. 1978, PASP, 267

\bibitem[{Wise {et~al.}(2018)Wise, {Dodson-Robinson}, Bevenour, \&
  Provini}]{Wise2018}
Wise, A.~W., {Dodson-Robinson}, S.~E., Bevenour, K., \& Provini, A. 2018, The
  Astronomical Journal, 156, 180

\end{thebibliography}

\onecolumn

\begin{appendix} 

  \section{RVs}

    Table \ref{tab:rvs} provides the full set of radial velocities and activity
    indicators from both the HARPS and ESPRESSO observations of \target. The
    columns are: 
    \begin{enumerate}
      \item modified barycentric Julian day (BJD);
      \item barycentric RV in $\kms$
      \item RV error in $\kms$
      \item CCF FWHM in $\kms$
      \item CCF bisector span in $\kms$
      \item activity index based on the Ca\,II H \& K lines (\ica)
      \item activity index based on the H$_\alpha$ line (\iha)
      \item activity index based on the Na\,I lines (\ina)
      \item instrument
    \end{enumerate}

      \begin{table*}[h] 
      
        \caption{Radial velocities and activity indicators (full table online-only).}
        \label{tab:rvs}
        \center
        \begin{tabular}{lllllllll}
        \hline
        \hline\noalign{\smallskip}
        Time & RV & $\sigma_\mathrm{RV}$ & FWHM & BIS & \ica & \iha & \ina & instrument \\ 
        BJD - 2\,400\,000 & [$\kms$] & [$\kms$] & [$\kms$] & [m s$^{-1}$] & -- & -- & -- & \\ 
        \hline\noalign{\smallskip}
        52943.85284 & 3.52687 & 0.00275 & 6.72630 & -0.03593 & 0.14124 & 0.18564 & 0.33982 & HARPS\\
        52989.71023 & 3.51946 & 0.00420 & 6.72352 & -0.02740 & 0.12679 & 0.19175 & 0.34004 & HARPS\\
        52998.68982 & 3.52675 & 0.00552 & 6.70572 & -0.03353 & 0.12621 & 0.18844 & 0.34867 & HARPS\\
        \ldots\\
        58489.58447 & 3.38982 & 0.00044 & 7.21972 & -0.09453 & 0.13210 & 0.18537 & 0.34246 & ESPRESSO\\
        58559.56959 & 3.38725 & 0.00058 & 7.22624 & -0.09247 & 0.11902 & 0.18658 & 0.34051 & ESPRESSO\\
        58559.58456 & 3.38740 & 0.00049 & 7.22581 & -0.09132 & 0.12758 & 0.18602 & 0.34091 & ESPRESSO\\
        \hline
      \end{tabular}
    \end{table*}

  \section{Analysis of the TESS light curve}
  \label{app:tess}

\begin{figure*}[h]
      \centering
      \includegraphics[width=\hsize]{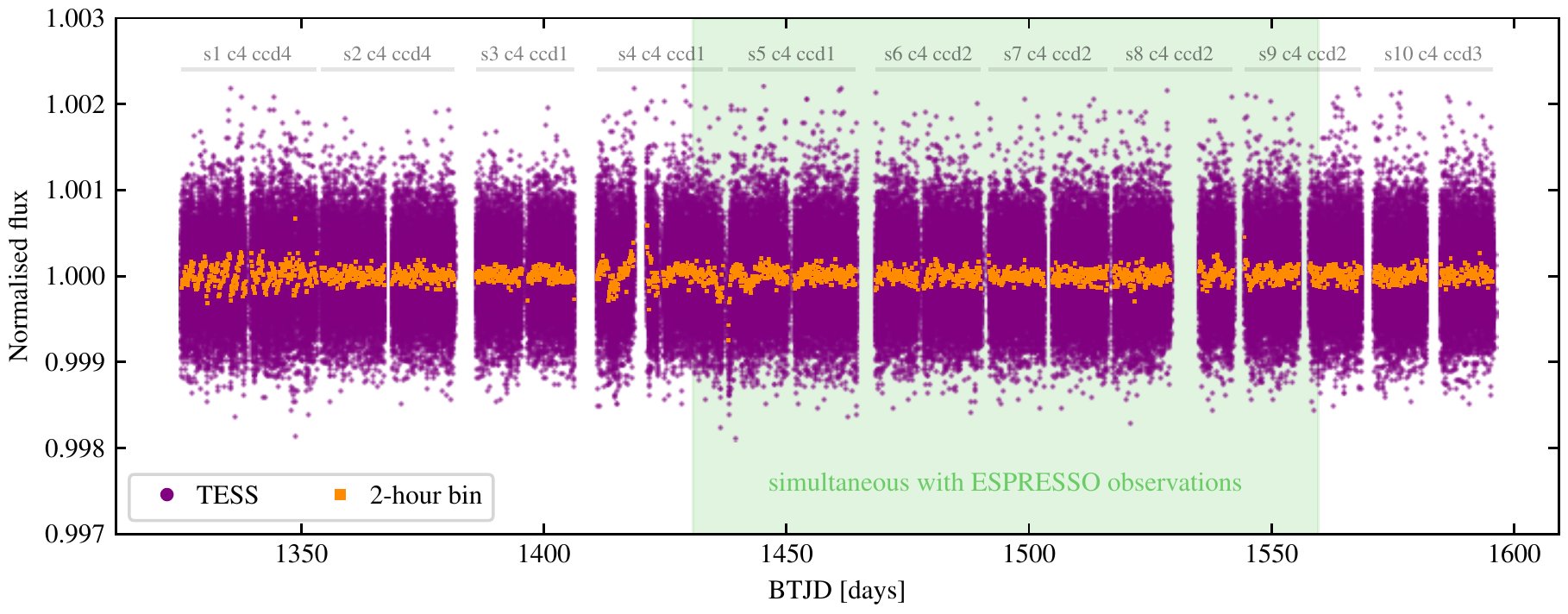}
      \caption{Merged \TESS light curve from the first ten sectors. The camera
        and CCD number with which \target was observed in each sector is
        indicated at the top, as well as the period of ESPRESSO observations.
        The orange points show the binned LC over a 2-hour window.}
      \label{fig:tess_lc}
    \end{figure*}

    The \TESS mission is set to observe \target during the full first year of
    its nominal two-year mission. Using the Lightkurve package
    \citep{lightkurve}, we downloaded and extracted the Pre-search Data
    Conditioning (\texttt{PDCSAP\_FLUX}) light curves (LC) produced by the
    Science Processing Operations Center from the Mikulski Archive for Space
    Telescopes (MAST%
    \footnote{
      \href{https://mast.stsci.edu/portal/Mashup/Clients/Mast/Portal.html}%
           {mast.stsci.edu/portal/Mashup/Clients/Mast/Portal}
    }%
    ). As of June 2018, data from the first ten sectors are available, with a
    baseline of 243 days. The individual LCs were then merged by adjusting the
    mean of the flux in each sector, and outliers were removed with a
    5-sigma-clipping procedure.
    This results in the merged LC shown in \fig{fig:tess_lc}, which also
    includes an indication of the period where TESS observations are
    simultaneous with ESPRESSO. 
    
    The merged LC shows a weighted rms of 431
ppm. Using the
    relations between active-region lifetime, spot size, and stellar effective
    temperature determined by \citet[][their Eq. 8]{Giles2017}. This leads to an
    estimate of 25.57
days for the decay lifetime of active regions
    in the stellar surface. This relation was built for star spots, since these
    have a larger effect in the brightness variations when compared with
    faculae. In the Sun, faculae tend to live longer than spots
    \citep{Solanki2006,Shapiro2017}.

    \subsection{Transit search}

    Without performing any additional detrending, we applied the Transit Least
    Squares (TLS) algorithm \citep{Hippke2019} on the merged LC. The TLS
    algorithm is optimised for detecting shallow transits by using a realistic
    transit shape with ingress and egress, as well as stellar limb darkening
    \citep{Mandel2002}, instead of the box-shaped function used in the more
    common Box Least Squares (BLS) algorithm \citep{Kovacs2002}. The TLS
    implementation is publicly available%
    \footnote{\href{http://github.com/hippke/tls}{github.com/hippke/tls}}%
    , requiring as inputs the stellar limb darkening, mass, and radius. We used
    the limb darkening estimates available in the TIC catalogue and the mass and
    radius in Table \ref{tab:parameters}. The output is the so-called signal
    detection efficiency (SDE) for the trial periods. According to
    \citet{Hippke2019}, SDE values larger than 7 result in a false-positive rate
    of 1\%.

    \begin{figure}[h]
      \centering
      \includegraphics[width=0.49\hsize]{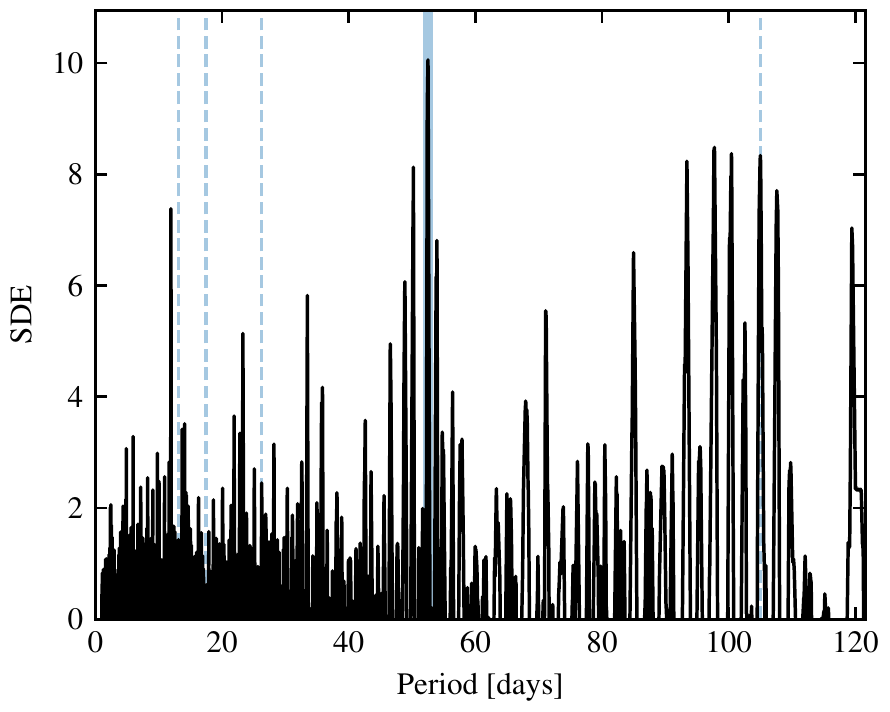}
      \includegraphics[width=0.49\hsize]{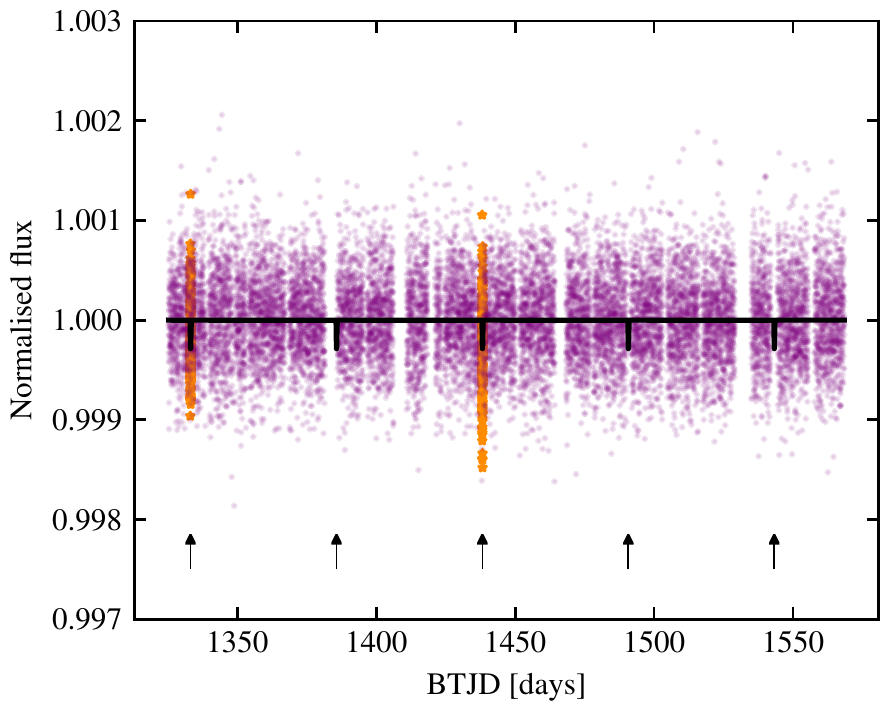}
      \caption{\emph{Left:} TLS periodogram for the TESS light curve of
      \target. Blue line represents the highest peak, at a period of 52.52
      days and with an SDE of 10.05. Dashed blue lines show harmonics of
      this period. \emph{Right:} Full \TESS light curve of \target with
      predicted transit curves (black) and in-transit data points (orange) for
      the highlighted period.}
      \label{fig:tls}
    \end{figure}

    This analysis results in the detection of a significant periodic signal at
    52.52 days, with a transit depth of 0.99971 and an SDE of 10.05 (see
    \fig{fig:tls}). However, 3 of the 5 transits are without data because they
    happen in the beginning or the end of gaps in the LC (mostly between \TESS
    sectors). A visual inspection of the LC together with the predicted transits
    confirms the occurrence of three transits during gaps (\fig{fig:tls}, right
    panel). One of the predicted transits happens at the start of sector five,
    where the data show a ramp up lasting about one day. Therefore, although the
    SDE of this signal should be considered significant, we believe that this is
    a false positive created by instrumental trends or the detrending procedures
    applied to the edges of each sector. 
    
    We then attempted to detrend the merged LC before repeating a transit
    search. The detrending aims to filter out as much noise as possible
    (outliers, stellar variability, residual instrumental signals) while
    preserving the transit signals. This step is often crucial to enhance the
    sensitivity of the transit search. For this reason, we tried two slightly
    different detrending approaches, both following the same scheme: we apply a
    low pass filter to obtain a smoothed version of the LC. The LC is then
    divided by the smoothed version and filtered for outliers using a
    5-sigma-clipping method. The difference between the two detrending methods
    is the type of low-pass filter used. First, we used a third order spline
    filter, with break points every 0.5 days, followed by a 5-sigma-clipping
    procedure \citep[for details, see][]{Barros2016}. For the second method, we
    used a Savitzky-Golay filter (as implemented in the \texttt{lightkurve}
    package) with a window size of 1.4 days and a second order polynomial
    detrending.

    These detrended LCs were then searched for transits with the BLS algorithm.
    The search was carried out over periods ranging from 0.5 to 97 days (60\% of the
    full duration of the LC). The highest signal in the BLS periodogram, after
    both detrending methods, is at a period of 28.4 days, but it has a low
    significance. Visual inspection allows us to discard the possibility of a
    transit. We conclude that there are no clear signs of transiting planets
    with periods up to 97 days.
    
    \subsection{Rotation period}

    We searched the \TESS LC for a periodic signal that can be associated to
    stellar rotation using four different methods: the GLS periodogram, the
    autocorrelation function \citep[ACF, e.g.][]{McQuillan2014}, the wavelet
    power spectra \citep[PS, e.g.][]{Torrence1998}, and the gradient of the
    power spectra (GPS\footnote{Not to be confused with Gaussian processes,
    GPs.}, \citealp{Shapiro2019}; Amazo-Gomez et al., in prep.)

    The GPS method in particular attempts to determine the rotation period from
    the enhanced profile of the high-frequency tail of the power spectrum by
    identifying the point where the gradient of the power spectrum reaches its
    maximum value. Such a point corresponds to the inflection point (IP), that is, a
    point where the concavity of the power spectrum changes sign.
    \citet{Shapiro2019} show that the period corresponding to the inflection
    point is connected to the stellar rotation period by a calibration factor
    equal to $\alpha_{\rm Sun} = 0.158$, for Sun-like stars.

\begin{figure*}
    \centering
    \includegraphics[width=16cm,clip]{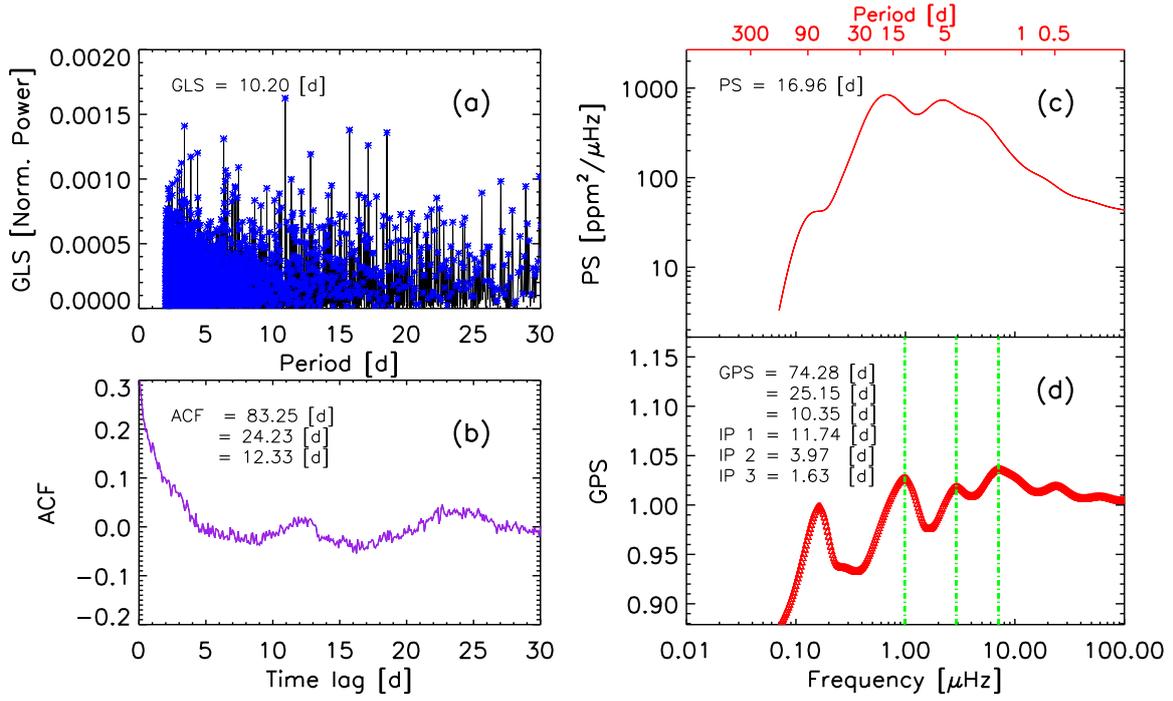}
    \caption{Results from the rotation period analysis showing the GLS
      periodogram (panel a),  ACF (panel b), power spectrum (panel c),
      and GPS (panel d) of the \TESS LC. Each panel displays the most
      prominent periods detected with each method.}
    \label{fig:tess_rot}
    \end{figure*}

    The results from the four methods are presented in \fig{fig:tess_rot} and
    can be summarised as follows: the GLS periodogram suggests a periodic signal
    of 10.2 days, but with a low relative power; the ACF shows periodic signals
    at 24.25 days and 12.34 days. The PS, in panel (c), shows two peaks at 16.96
    days and 6.15 days. The GPS method shows three enhanced inflection points
    with enough amplitude to determine three different periodicities. The
    inflection points at 11.74, 3.97, and 1.63 days correspond to periodic
    signals at 74.28, 25.15, and 10.35 days after applying the calibration
    factor $\alpha_{Sun}$.

    From the values obtained using the four different methods, we can see that
    both the GLS and GPS methods detect a periodicity close to 10 days. The
    strongest signal in the ACF is around 24.25 days, in agreement with the
    second enhanced signal from GPS, of 25.15 days. The values obtained with the
    ACF and GPS are close to those obtained from spectroscopy ($v \sin i$) and
    with the periodicities seen in some activity indicators, suggesting a
    stellar rotation period for \target of about 25 days.

\end{appendix}

\end{document}